\documentclass[preprint,12pt]{elsarticle}

\usepackage{amssymb}
\usepackage{bm}

\journal{Nuclear Physics A}

\newcommand{\nuc}[2]{$^{#1}$\textrm{#2}}

\newcommand{\text}[1]{\mbox{\scriptsize{#1}}}
\newcommand{\disregard}[1]{}

\newcommand{\AnA}[1]{#1}

\newcommand{\MiB}[1]{#1}

\usepackage{CJKutf8}
\usepackage{color}

\begin{document}

\begin{CJK*}{UTF8}{gbsn}
\begin{frontmatter}



\title{Properties of nuclei in the nobelium region studied within the covariant, Skyrme, and Gogny
energy density functionals}


\author{J. Dobaczewski$^{1-4}$, A.V. Afanasjev$^5$, M. Bender$^{6,7}$, L.M. Robledo$^8$, and Yue Shi$^{9}$ (石跃)
}

\address{
$^1$Department of Physics, University of York, Heslington, York YO10 5DD, United Kingdom\\
$^2$Department of Physics, P.O. Box 35 (YFL), University of Jyv\"askyl\"a\\ FI-40014  Jyv\"askyl\"a, Finland\\
$^3$Institute of Theoretical Physics, Faculty of Physics, University of Warsaw\\ Pasteura 5, PL-02-093 Warsaw, Poland\\
$^4$Helsinki Institute of Physics, P.O. Box 64, FI-00014 Helsinki, Finland\\
$^5$Department of Physics and Astronomy, Mississippi State University\\ Mississippi State, Mississippi 39762, USA\\
$^6$Universit{\'e} de Bordeaux, Centre d'Etudes Nucl{\'e}aires de Bordeaux Gradignan\\ UMR5797, F-33175 Gradignan, France\\
$^7$CNRS/IN2P3, Centre d'Etudes Nucl{\'e}aires de Bordeaux Gradignan\\ UMR5797, F-33175 Gradignan, France\\
$^8$Departamento de F\'{\i}sica Te\'orica, Universidad Aut\'onoma de Madrid\\ E-28049 Madrid, Spain\\
$^9$National Superconducting Cyclotron Laboratory, Michigan State University, East Lansing, Michigan, 48824-1321, USA\\
}

\begin{abstract}

We calculate properties of the ground and excited states of nuclei in
the nobelium region for proton and neutron numbers of $92\leq
Z\leq104$ and $144\leq N\leq156$, respectively. We use three
different energy-density-functional (EDF) approaches, based on
covariant, Skyrme, and Gogny functionals, each with two different
parameter sets. A comparative analysis of the results obtained for
quasiparticle spectra, odd-even and two-particle mass staggering,
and moments of inertia allows us to identify single-particle and
shell effects that are characteristic to these different models and
to illustrate possible systematic uncertainties related to using the
EDF modelling.

\end{abstract}


\end{frontmatter}
\end{CJK*}

\newpage

\section{Introduction}
\label{sec1}

Recent experimental studies of nuclei in the nobelium region provided
rich spectroscopic data~\cite{[Her08],[The15]}, which, in
principle, can be used as a benchmark information for extrapolations into the
region of superheavy nuclei. Numerous theoretical studies are aimed at
modelling of these spectroscopic
data~\cite{[Cha77],[Egi00b],[Sob01],[Dug01w],[Ben03d],[Afa03],[Vre05],[Del06],[Sob07],%
[Sob11],AS.11,[Zha12],[Lit12],[Liu12],[War12],[Pra12],[Zha13],[Afa13],[Shi14a]}
so as to make such extrapolations as reliable as possible.

The estimation of theoretical uncertainties is one of the most
essential aspects of extrapolating nuclear models into exotic
nuclei~\cite{[Dob14a]}. One, fairly easy part of it, is the
evaluation of statistical uncertainties of observables that are
related to the uncertainties of model parameters adjusted, in one way or
another, to experimental data. Another one, very difficult, pertains
to those systematic uncertainties related to the definition and contents of the
different terms that make up the models
themselves. An obvious strategy, which, anyhow, gives us only a
limited glimpse on possible systematic uncertainties, is to study a
set of variants of a given model, and to analyze differences obtained for calculated
observables.

In the present study, we aim at such an analysis of results obtained
within three fairly different energy-density-functional (EDF)
approaches. Namely, we employ the covariant EDFs~\cite{[Vre05]}, with one classic
(NL1~\cite{NL1}) and one recent (NL3*~\cite{NL3s}) parameter set,
Skyrme EDFs~\cite{[Ben03]}, with one classic (SLy4~\cite{[Cha98a]}) and
one recent (UNEDF2~\cite{[Kor14]}) parameter set, as well as Gogny
EDFs \cite{[Per15a]}, again with
one classic (D1S~\cite{[Ber91b]}) and one recent (D1M~\cite{[Gor09a]})
parameter set.

Our goal is thus to determine, present, and compare results obtained
within these six models for a common set of calculated observables.
We aim at performing these analyses within the most similar and/or
equivalent conditions, so as to meaningfully discuss general {\em
qualitative} similarities and differences.
%
%
\MiB{In all cases, pairing correlations are treated on the Bogoliubov level. In
the literature, it is customary to label such calculations as Hartree-Fock-Bogoliubov (HFB)
in the non-relativistic cases and as relativistic Hartree-Bogoliubov (RHB) for the specific
variant of relativistic mean-field model used here, thereby emphasizing that HFB-like equations
are solved instead of simpler HF+BCS equations.
Only in the calculations with a Gogny force, however, the same effective interaction
is used to determine direct, exchange, and pairing matrix elements. By contrast, in case of
Skyrme EDF and the relativistic approach particle-hole and pairing matrix elements relate
to different effective interactions, which can be used to simplify their
form and phenomenological adjustment. Also, in the RHB approach all exchange terms are neglected,
whereas in the case of many Skyrme parameterizations some are modified in order to
improve the description of data \cite{[Ben03]}.
None of these formal differences is relevant for our discussion, and we will use whenever
possible the generic notion of an EDF method for all three approaches. Indeed, in all
cases the total energy can be cast into the form of a functional of normal and anomalous
one-body density matrices from which the equations-of-motion are then derived by variation.
}
We do not
attribute too much of an importance to {\em quantitative}
similarities and differences between the obtained results, especially
when the models are compared to experimental data. Indeed, a detailed
agreement with the  data may crucially depend on specific model-parameter
adjustments, or on various corrections taken into account or
disregarded. The {\it phenomenological} EDF used include a limited set
of parameters (typically between seven and twenty) and aim at a global
description of a wide variety of nuclear properties and therefore a
perfect agreement with experimental data is out of reach at present.
Certainly, in the future all EDF approaches will be
improved; here we only look into generic properties obtained for
selected current global parameterisations thereof.

In the present analysis, we systematically calculated the ground
states of even-even and odd-mass nuclei from uranium ($Z=92$) to
rutherfordium ($Z=104$) and for neutron numbers between $N=144$ and
156. \AnA{The selection of this region of heaviest actinides/lightest
superheavy nuclei is guided by the need for reliable experimental
data on spectroscopic properties (in particular, on the single-particle
energies of deformed one-quasiparticle states) based on which the
extrapolability of a given theory/functional towards region of
superheavy nuclei may be judged.} In addition, we determined low-lying
quasiparticle spectra of
odd-mass nuclei and low-spin moments of inertia of even-even nuclei.
The main thrust of the analysis was on the attempt to identify
single-particle and shell-structure properties of these nuclei by
looking at many-body observables such as masses, odd-even and two-particle mass
staggering, and excitation energies.

The paper is organised as follows. Selected theoretical aspects of
our calculations are presented in section~\ref{sec2}, with four
subsections discussing the methods related to obtaining results for the
Skyrme EDF SLy4 (\ref{sec2b}), Skyrme EDF UNEDF2 (\ref{sec2d}), Gogny
EDFs (\ref{sec2c}), and covariant EDFs (\ref{sec2a}). The results of
the calculations are given in section~\ref{sec3}, with subsections
devoted to the Nilsson diagrams (\ref{sec3b}), quasiparticle spectra
(\ref{sec3a} and \ref{sec3c}), odd-even and two-particle mass staggering
(\ref{sec3d}), and moments of inertia (\ref{sec3e}).
Conclusions are presented in section~\ref{sec4}.

\section{Calculation details}
\label{sec2}

\subsection{Skyrme energy density functional SLy4}
\label{sec2b}

For the Skyrme EDF SLy4 \cite{[Cha98a]},
the calculations were carried out with the Skyrme HFB solver \textsc{CR8}
whose development over the years has been documented in
Refs.~\cite{[Bon87],[Gal94],[Ter95],hellemans12}.
It uses a 3D coordinate space
mesh representation of single-particle states along the lines of the
solver \textsc{EV8} described in Ref.~\cite{[Rys15]}, but is
extended in such a way that intrinsic time-reversal invariance can be
broken and that HFB equations are solved instead of HF+BCS.
Single-particle states are represented in a cubic box
of $32^3$\,fm$^3$ with a step size of 0.8\,fm between
discretization points. Imposing triaxial symmetry, only 1/8 of the
box has to be represented numerically,
meaning that only a $20 \times 20 \times 20$ mesh is to
be treated.

When calculated with SLy4,
the ground states of even-even nuclei
considered here are all axial, and the blocked states of odd-$A$
nuclei also remain almost axial.
All blocked calculations were initialized with the ground states of
adjacent even-even nuclei.
Self-consistent blocking was performed by considering the
quasiparticle state dominated by a given eigenstate of the
single-particle Hamiltonian and by exchanging the corresponding columns of the HFB $U$
and $V$ matrices after the diagonalization of the HFB Hamiltonian,
which in turn was constructed using the mean fields of the blocked
solution from the previous iteration, see, e.g.,
Refs.~\cite{[Gal94],[Ber09],[Sch10]}.

To avoid mixing of quasiparticle states
with different average values of the angular momentum component $\langle \hat{j}_\parallel \rangle$ parallel to the symmetry axis of the initial configuration in the
diagonalization of the HFB Hamiltonian, the many-body expectation
value of $\langle \hat{J}_\parallel \rangle$ was held fixed with a cranking
constraint at the value of $\langle \hat{J}_\parallel \rangle$ equal to the one of
the blocked quasiparticle state. As the code \textsc{CR8} allows for
triaxial shapes the mixing cannot be fully suppressed. As a consequence, the
blocked HFB states are not necessarily orthogonal even when they have
different average value of $\langle \hat{J}_\parallel \rangle$.

In this respect,
blocking the $\langle \hat{j}_\parallel \rangle = 1/2$ levels presents a particular difficulty. Without a cranking constraint, the code \textsc{CR8}
very often converges toward a solution where the many-body expectation
value $\langle \hat{J}_\parallel \rangle$ is close to zero, and where the blocked
quasiparticle in the spectrum of eigenstates of the HFB Hamiltonian
is mixed with other low-lying quasiparticles of different
$\langle \hat{j}_\parallel \rangle$. For these states, using or not using the
cranking constraint might in some cases
make a difference of the order of 100 to 200\,keV.
An example is the ground state of $^{251}$Cf. In blocked calculations with
cranking constraint, there is a low lying $\langle \hat{J}_\parallel \rangle^\pi = 1/2^+$
level at 45\,keV excitation energy above the calculated $\langle \hat{J}_\parallel \rangle^\pi = 3/2^+$ ground state.
In the calculations without cranking constraint, the energy of the $3/2^+$ state
does not change much, but the $1/2^+$ level is lowered by about 180\,keV and
becomes the ground state, in agreement with experiment, but at the expense
of the blocked quasiparticle being a strong mixture of
$\langle \hat{j}_\parallel \rangle = 1/2$ and $\langle \hat{j}_\parallel \rangle = 3/2$, and
of having an angular momentum $\langle \hat{J}_\parallel \rangle$ that cannot be easily interpreted within the
strong coupling model anymore.
For blocked states with higher $\langle \hat{j}_\parallel \rangle$, the
mixing is always much smaller.

For the coupling constants of the so-called time-odd terms that contribute to cranked and blocked states, the same "hybrid" choice
was made as in many earlier calculations \cite{[Ben03d],[Bon87],[Gal94],[Ter95],hellemans13,duguet01,chatillon06,chatillon07,ketelhut09,bender12,[Cwi99w]}.
They were set to the "native" values
dictated by a density-dependent Skyrme two-body force for all terms except for
those that multiply terms that couple two derivatives and two Pauli spin matrices. The latter
were set to zero for reasons of Galilean invariance and internal consistency, cf.\ Refs.~\cite{[Bon87],hellemans13} for further discussion.

In the present study, neutron and proton surface pairing interactions were used, with the strengths
adjusted to the three-point gaps $\Delta^{(3)}_n$ of \nuc{247}{Cf} ($N=149$) and $\Delta^{(3)}_p$
of \nuc{249}{Bk} ($Z=97$),
\MiB{as defined by Eqns.~(\ref{D3N}) and~(\ref{D3P}) below,
}
leading to
\begin{eqnarray}
V_{n} & = & -1240 \; \mbox{MeV} \; \mbox{fm}^{3}, \\
V_{p} & = & -1575 \; \mbox{MeV} \; \mbox{fm}^{3}.
\end{eqnarray}
The pairing-active spaces were limited by the soft cutoffs of 5\,MeV above
and below the neutron and proton Fermi energies as described in
Ref.~\cite{rigollet99}.
With that, the proton pairing strength is significantly larger than that given by the
standard value of $V_n = V_p = -1250$\,MeV\,fm$^{3}$
used for heavy nuclei in previous publications
\cite{duguet01,[Ben03d],chatillon06,chatillon07,ketelhut09,bender12}
(which all have been carried out using the Lipkin-Nogami (LN) scheme, though). These previous values were
obtained by adjusting moments
of inertia in rotational bands of heavy nuclei \cite{rigollet99}.
There is, however, also some published work on very heavy nuclei \cite{[Cwi99w],cwiok05}
where the pairing interaction of volume type was adjusted to some
$\Delta^{(3)}_q$ values of odd-$A$ actinide nuclei.

In principle, the like-particle pairing interaction should be of
isovector type, which implies $V_{n} = V_{p}$.
Differences between the adjusted neutron and proton pairing strengths can have many
reasons:
(i) the compensation of imperfections of the calculated single-particle spectra,
(ii) the compensations of the cutoff energy that is chosen to be the same in neutron and proton phase spaces
(iii) the compensation of the imperfections of the chosen form of the pairing
interaction itself.
Note that if there were none of the above mentioned deficiencies,
the proton pairing strength would have to
be \textit{smaller} than the neutron pairing strength, so as to
compensate for the absence of Coulomb pairing in our calculations.

Our attempts to adjust the neutron pairing strength to
$\Delta^{(3)}_n$ in \nuc{251}{Cf} ($N=153$), as done in
Ref.~\cite{[Shi14a]} and in the present work for UNEDF2, led to the values of $V_{n} = -1060$
and $V_{p} = -1565$\,MeV\,fm$^{3}$. With such much weaker neutron
pairing strength, in ground states of many of the lighter odd-$N$
nuclei, the pairing disappeared. This seems to be connected to an anomaly
of the calculated single-particle spectra of heavy $N=153$ isotones
that translates into a much larger values of $\Delta^{(3)}_n$ for
these nuclei than for the neighbouring ones.

\subsection{Skyrme energy density functional UNEDF2}
\label{sec2d}

For the Skyrme EDF UNEDF2~\cite{[Kor14]}, the calculations were performed using
the symmetry-unrestricted code \textsc{hfodd} (v268h)~\cite{[Sch12]}. The HFB equations
were solved by expanding single-particle wave functions on 680
deformed harmonic-oscillator (HO) basis states, with HO frequencies of $\hbar
\Omega_x = \hbar \Omega_y = 8.4826549$ MeV and $\hbar \Omega_z = 6.4653456$
MeV. This corresponds to including the HO basis states up to $N_x=N_y=13$ and
$N_z=14$.

The time-odd coupling constants of the Skyrme EDF UNEDF2 were determined by
the local-gauge-invariance arguments, as defined in Ref.~\cite{[Ben02]}. The
HFB equation adopted a quasiparticle cutoff energy of 60\,MeV. Within this model space, the neutron and proton
pairing strengths were readjusted to match the experimental values of $\Delta^{(3)}_n$ and
$\Delta^{(3)}_p$ in $^{251}$Cf and $^{249}$Bk, respectively. This resulted in values of
\begin{eqnarray}
V_{n} & = & -233.889 \; \mbox{MeV} \; \mbox{fm}^{3}, \\
V_{p} & = & -280.330 \; \mbox{MeV} \; \mbox{fm}^{3}
\end{eqnarray}
that are larger by about 25 and 50\,MeV fm$^{3}$, respectively, than those corresponding to the original UNEDF2 values, where
the LN pairing corrections were included, though.

\subsection{Gogny energy density functionals}
\label{sec2c}

Odd mass nuclei were described using the HFB method
with full blocking and taking into account all possible time-odd fields
coming from the potential part of the Gogny interaction. The original
formulation of the density-dependent part of the interaction is preserved
and therefore no additional time-odd density-dependent terms are added
to it. This assumption has to be verified in the future as there are
indications that such kind of terms could be required to recover basic
properties of odd-odd systems \cite{rob14}.

The computer code \textsc{atb} was used in the calculations
\cite{rob12}. Axial symmetry was preserved in the calculations but
reflection symmetry was not. Therefore, the projection of the angular
momentum along the intrinsic axial-symmetry axis, $\Omega=\langle \hat{j}_\parallel \rangle$, is a good
quantum number, but this is not necessarily the case for parity.
However, many of the states analyzed show a mean value of the parity
operator, which is rather close to either plus or minus one, allowing
the labelling of those states with a definite parity. As $\Omega$ is a good
quantum number, states with different values of $\Omega$ are automatically
orthogonal. States with the same $\Omega$ value, same parity and similar
deformation parameters are imposed to be orthogonal to each other
by using the traditional technique of constraints. The way the constraints
are handled does not depend on the nature of the constrained operator
as only its mean value and gradient are required \cite{rob11}.
The code  expands the quasiparticle operators in a
harmonic oscillator basis with 15 shells ($680\times2$ states)
although axial symmetry reduces the maximum size of matrices to
$120\times2$, corresponding to the $\Omega=1/2$ block.

In the odd-mass (odd number parity)
case, the HFB iterative minimization process requires a starting mean-field
state with given characteristics. An HFB mean-field state with even number parity (obtained with a time-even
code) was used to generate this starting configuration.
The mean values of neutron $N$ and proton $Z$ number
operators were constrained to the required values for the odd nucleus under
consideration. To avoid spurious good-parity solutions,
which could result from the propagation of self-consistent symmetries
in the HFB method, a small octupole moment was also constrained.

Ten quasiparticle states with the lowest one-quasiparticle
energies coming out of this calculation were considered as starting
configurations for the time-odd code \textsc{atb}. They were obtained with
the standard ``blocking procedure" consisting of swapping the appropriate
columns in the $U$ and $V$ matrices of the Bogoliubov transformation. Each
of the starting configurations was labelled with the corresponding $\Omega$ quantum
number, which was preserved all along the minimization process. The first starting
configuration with a given value of $\Omega$ usually converges to the
lowest-energy band head with angular momentum $I=\Omega$. For the same
value of $\Omega$, the second starting
configuration usually converges to the same first solution, unless an
orthogonality constraint to that state is imposed
(including both the $\Omega$ and $-\Omega$ states). For the third starting
configuration for a given $\Omega$ value, the orthogonality constraints with respect to
the first and second states are required. In principle, for the $n$-th
starting configuration $2(n-1)$ orthogonality constraints to the
states previously obtained are required. Although the number of
constraints can be relatively high, this is not a formal problem for
the gradient method used in the code, as handling of
constraints is very simple to implement \cite{rob11}. The
orthogonality constraint is also important to ensure the
orthogonality of the physical states. If such orthogonality is not
imposed, even if the states are different, the relative excitation
energies can be shifted. In the present calculations only two
starting configurations for each value of $\Omega$ were considered.

For the calculation of the properties of even-even nuclei required in
the evaluation of the pairing gap parameters, the code \textsc{HFBaxial} was
used. It has the capability to break reflection symmetry although the
phenomenon was not relevant in the present calculation.
Moments of inertia were determined using cranked HFB wave functions obtained
with the code \textsc{HFBtri}. In this code, the axial and time-reversal symmetries
were allowed to be broken in the minimization
process.

\subsection{Covariant energy density functionals}
\label{sec2a}

The RHB equations \cite{CRHB,[Afa03]}
were solved in the basis of an
anisotropic three-dimensional harmonic oscillator in Cartesian
coordinates. For all nuclei and states determined in this work, the
same basis deformation of $\beta_0=0.3$, $\gamma=0^{\circ}$ and
oscillator frequency of $\hbar \omega_0=41$A$^{-1/3}$\,MeV have been used. All
fermionic and bosonic states belonging to the shells up to $N_F=14$
and $N_B=20$ were taken into account when performing
diagonalization of the Dirac equation and matrix inversion of
the Klein-Gordon equations, respectively. As follows from detailed
analysis of Refs.\ \cite{[Afa03],AS.11}, this truncation of basis
provides sufficient accuracy of the calculations.

As the effective interaction in the particle-particle ($pp$) channel,
the central part of the non-relativistic Gogny
finite-range interaction
was used. The clear advantage of such a pairing force is that it
provides an automatic cutoff of high-momentum components. The motivation
for such an approach to the description of pairing was given in
Ref.\ \cite{CRHB}. The D1S parametrization of the Gogny force
was used here. No specific adjustment of its strength was used,
because it provided a reasonable description of the pairing indicators in
$^{249}$Bk and $^{249,251}$Cf and moments of inertia in $^{252,254}$No
\cite{[Afa03]}.

Two different covariant EDFs, namely, NL1
\cite{NL1} -- fitted to the nuclei in the valley of beta-stability,
and NL3* \cite{NL3s} -- tailored towards the description of
neutron-rich nuclei, were used in the current study. The covariant EDF NL1
was extensively used in the calculations of rotational
bands across the nuclear chart (see Ref.\ \cite{[Afa13]}).
Covariant EDF NL3* was less tested than NL1 with respect to the
description of rotating nuclei. However, its global performance is
well established \cite{AARR.14}.  Note that so far only these two
covariant EDFs were systematically confronted with experimental data on
single-particle states. For example, the study of predominantly
single-particle states in odd-mass nuclei neighbouring to the doubly
magic spherical nuclei was performed in Ref.\ \cite{LA.11}
within the relativistic particle-vibration coupling model employing
covariant EDF NL3*. In Ref.\ \cite{AS.11}, the first systematic study of the single-particle spectra
in deformed nuclei in rare-earth region and actinides was
performed with the covariant EDFs NL1 and NL3*.

It is interesting that the overall accuracy of the description of the
energies of deformed one-quasiparticle states \cite{AS.11} is slightly better in
the old covariant EDF NL1 than in the recent functional NL3*.
This suggests that the inclusion of extra
information on neutron rich nuclei into the fit of the functional NL3*
may lead to some degradation of the description of
single-particle states along the valley of beta-stability.
Note that these two functionals well reproduce deformation
properties of ground states of even-even actinides \cite{[Afa03],[Afa13]}
and indicate that they are axially symmetric.

A proper description of odd or rotating nuclei implies breaking of the time-reversal
symmetry of the mean field, which is induced by the unpaired nucleon
\cite{AA.10} or rotation \cite{TO-rot}. As a consequence, time-odd
mean fields and nucleonic currents, which cause the {\it nuclear magnetism}
\cite{KR.89} have to be taken into account. In the covariant EDF,
time-odd mean fields are defined through the Lorentz invariance, and
thus they do not require additional coupling constants.

The effects of blocking due to the odd particle were
included in a fully self-consistent way. This was done within the code \textsc{CRHB},
according to Refs.\ \cite{RBM.70,EMR.80,[Rin80]}. The blocked
orbital was specified by different tags such as
(i) dominant main oscillator quantum number $N$ of the wave
function,
(ii) dominant $\Omega$ quantum number of the wave
function,
(iii) particle or hole nature of the blocked orbital,
and (iv) position of the state within the specific
parity/signature/dominant-$N$/dominant-$\Omega$.
For a given odd-mass nucleus, possible blocked configurations were
defined from the analysis of calculated quasiparticle spectra in
neighboring even-even nuclei and the occupation probabilities of
the single-particle orbitals of interest in these nuclei.

Note that in the cases when the calculations of odd-mass nuclei
were performed only for the definition of the $\Delta^{(3)}$ indicators
(see Sec.\ \ref{sec3d} below), we restricted the analysis to 5--6
one-quasiparticle configurations with expected lowest total
energies, so as to properly determine the ground state of an
odd-mass nucleus. The calculations confirmed the conclusion of the
statistical analysis of Ref.\ \cite{AS.11} that absolute
majority of the one-quasiparticle configurations are axially
symmetric. However, some degree of triaxiality was obtained in
the $\nu 3/2[622]$, $\nu 1/2[501]$ and $\pi 1/2[400]$
configurations.

\section{Results}
\label{sec3}

In this section, we present the results obtained and we look for
signatures of shell effects in the systematics of many-body
observables. This is in addition to analyzing the eigenvalues of
mean-field Hamiltonians as a function of deformation, that is, the
Nilsson diagrams, which are not observables, but provide a useful
illustration of the underlying single-particle structure.

We note here that the calculation of many-body observables implies a
full self-consistency reached for every individual state, that is,
for ground and excited states. For example, quasiparticle spectra,
which we discuss below, always result from calculating differences of
total energies, determined separately for different many-body
self-consistent solutions. Because of that, each blocked HFB state
may have a slightly different quad\-ru\-pole, hexadecapole, or higher
deformation (see, for example, the results of statistical analysis
in Ref.\ \cite{AS.11}), which then feeds back to the mean and pairing fields. In
this way, the deformation-polarization effects, exerted by
one-quasiparticle states, are fully taken into account.

In odd-mass nuclei, quasiparticle excitations were obtained by blocking the
relevant levels when performing the HFB calculations. The spectra were
obtained by comparing the total energies of different configurations. The
procedure closely followed that of Refs.~\cite{[Afa03],[Sch10]}.

\subsection{Nilsson diagrams}
\label{sec3b}

\begin{figure}
\centerline{\includegraphics[width=\textwidth]{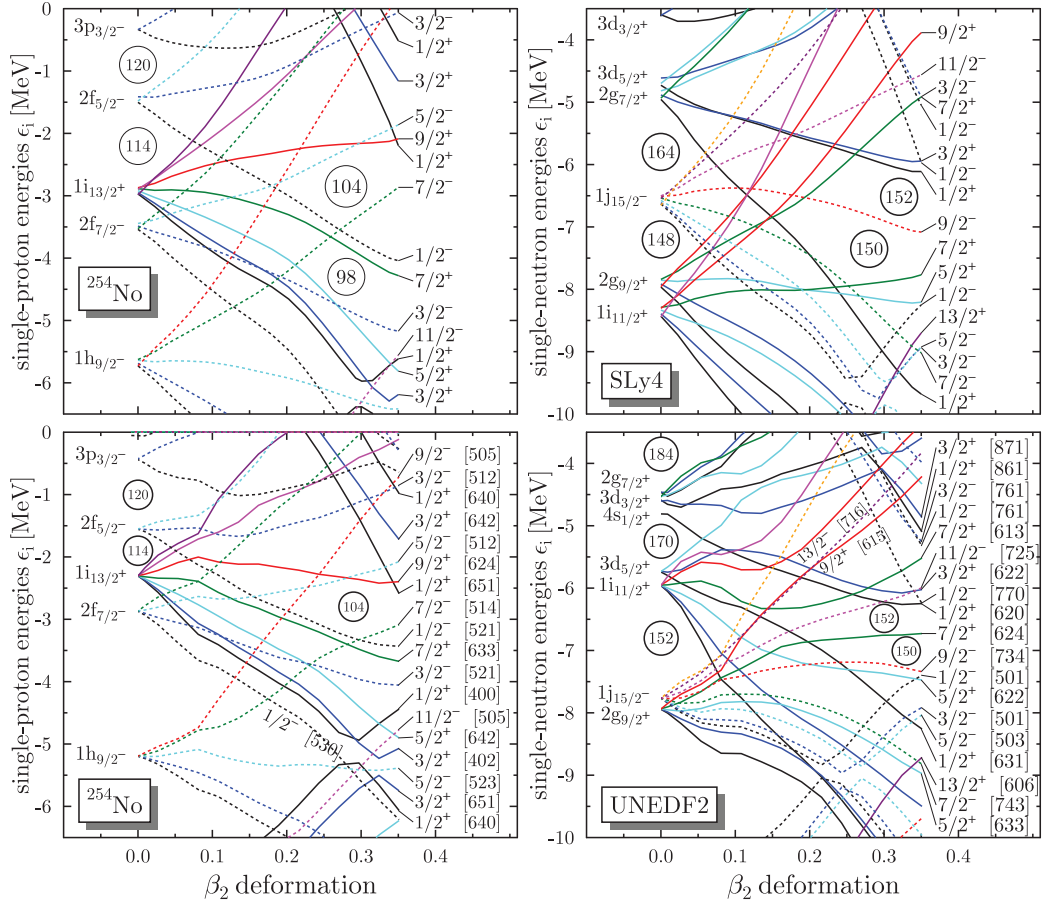}}
\caption{
\label{no254.sly4.nilsson2}
Proton (left panels) and neutron (right panels) Nilsson diagrams of \nuc{254}{No} obtained for the Skyrme EDF
SLy4 (upper panels) and UNEDF2 (lower panels). At spherical shapes, the orbitals are labelled with spherical
quantum numbers. For SLy4, at large
deformations, the deformed single-particle orbitals are labelled by the expectation values
$\langle \hat{j}_\parallel \rangle$ of the projection of the angular momentum on
the axial-symmetry axis. For UNEDF2, these orbitals are labelled by the Nilsson labels
$\Omega[Nn_z\Lambda]$ determined using code {\sc hfodd}. Solid and dashed lines
are used for the positive and negative parity states, respectively.
}
\end{figure}

\begin{figure}
 \centerline{\includegraphics[width=\textwidth]{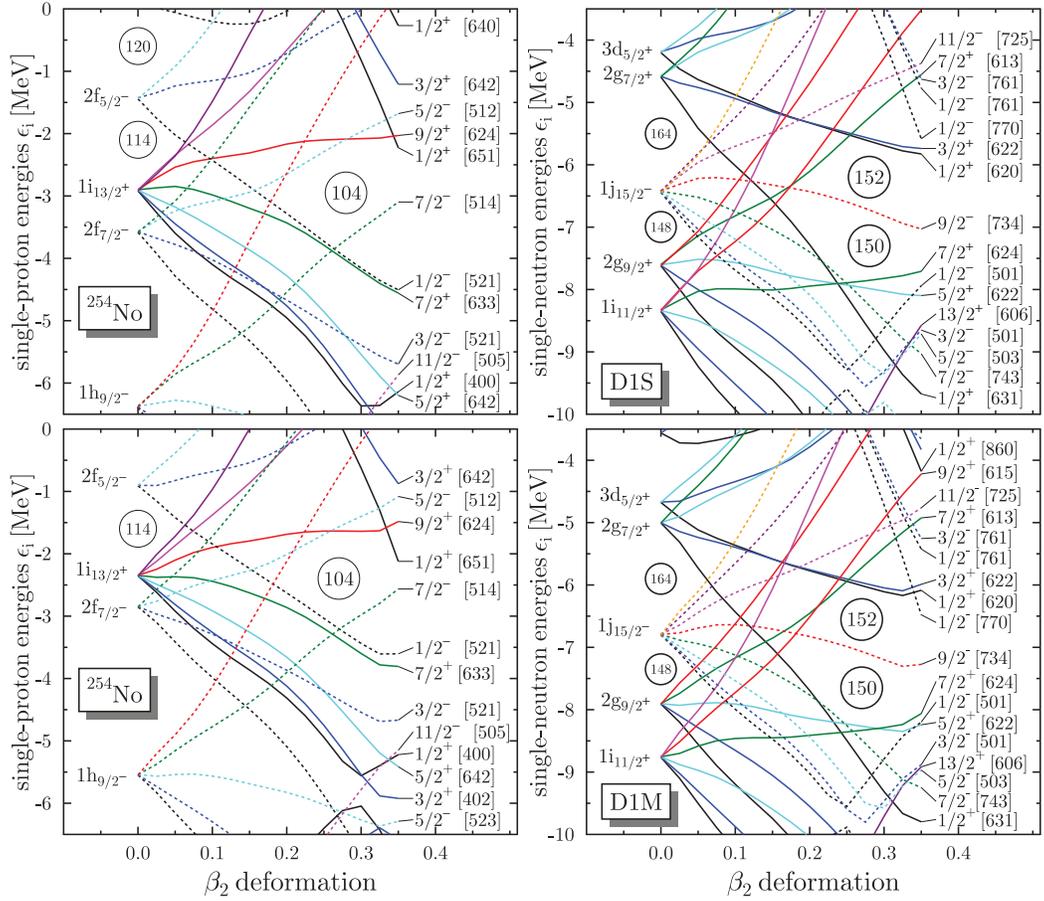}}
\caption{
\label{no254.d1s.nilsson2}
Same as in Fig.~\protect\ref{no254.sly4.nilsson2} but for the Gogny EDF D1S and D1M.
All results and dominant Nilsson labels were determined using the axial-symmetry code \textsc{HFBaxial}.
}
\end{figure}

\begin{figure}
\centerline{\includegraphics[width=0.49\textwidth,angle=0]{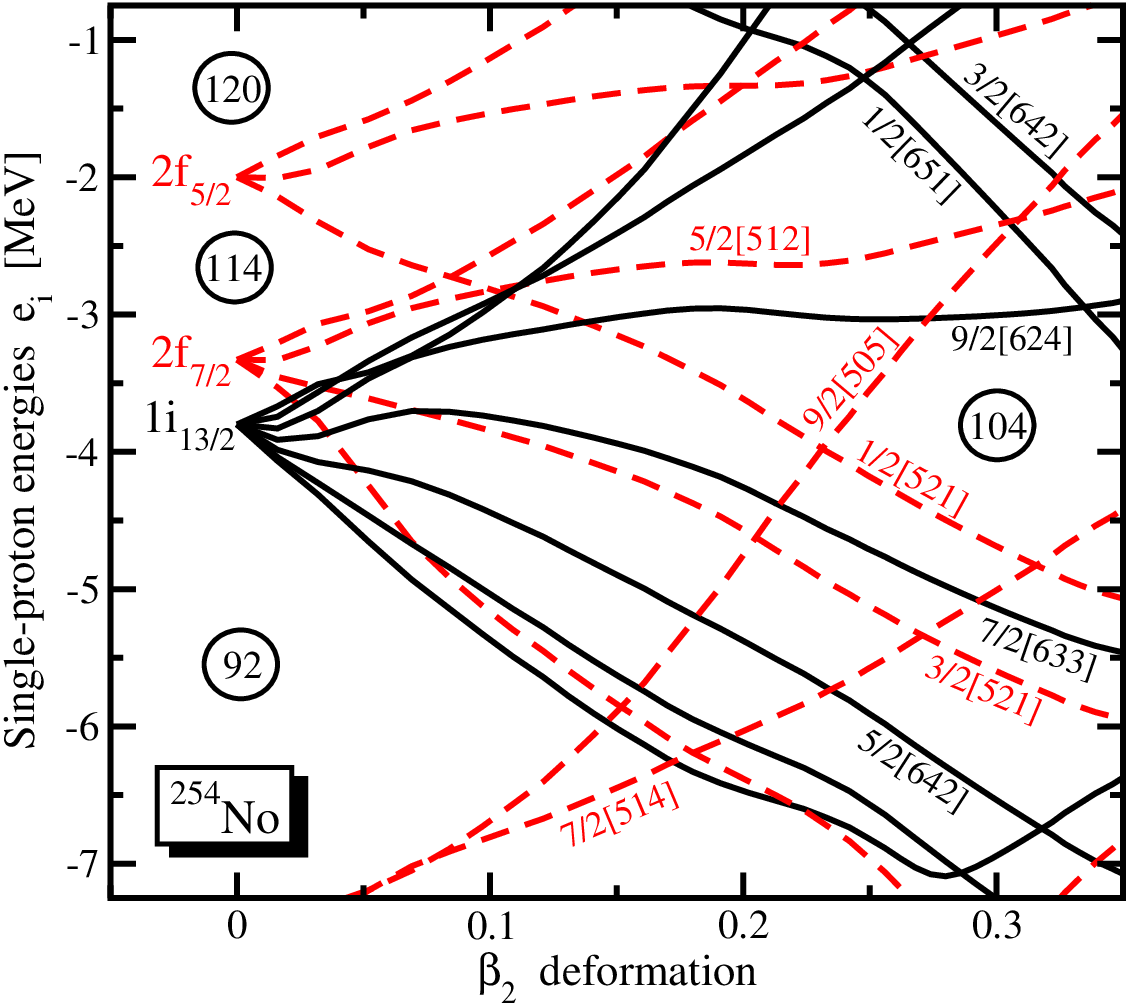}
            \includegraphics[width=0.49\textwidth,angle=0]{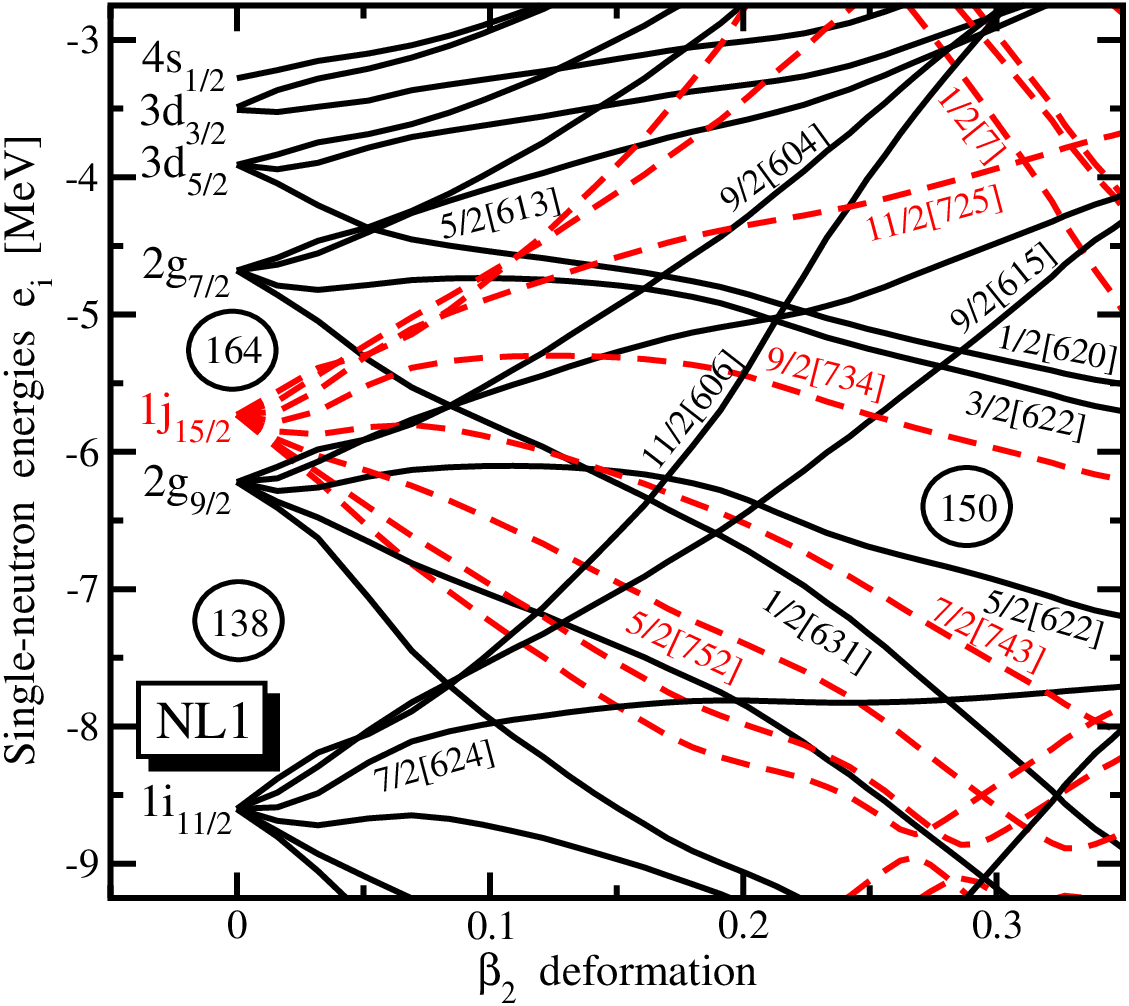}}
\centerline{\includegraphics[width=0.49\textwidth,angle=0]{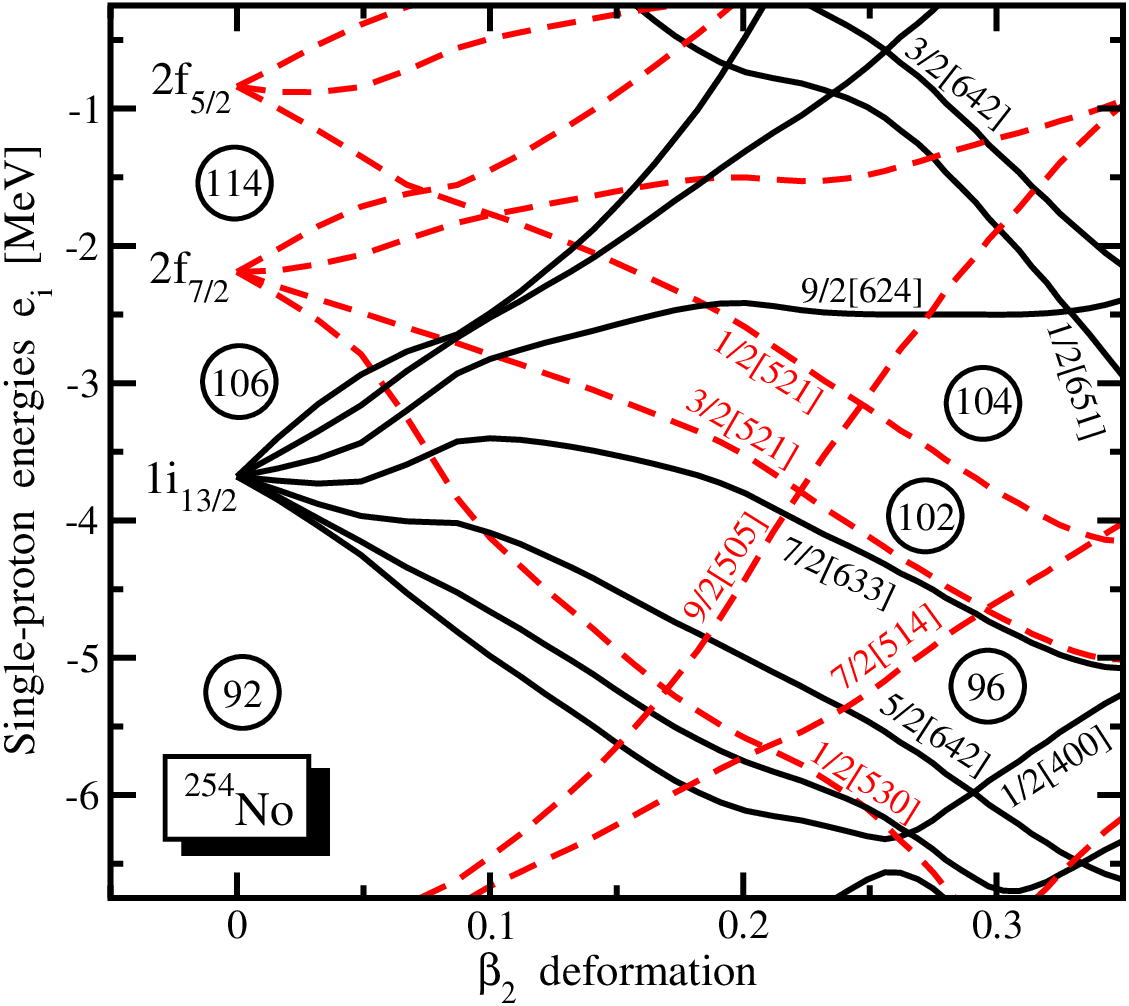}
            \includegraphics[width=0.49\textwidth,angle=0]{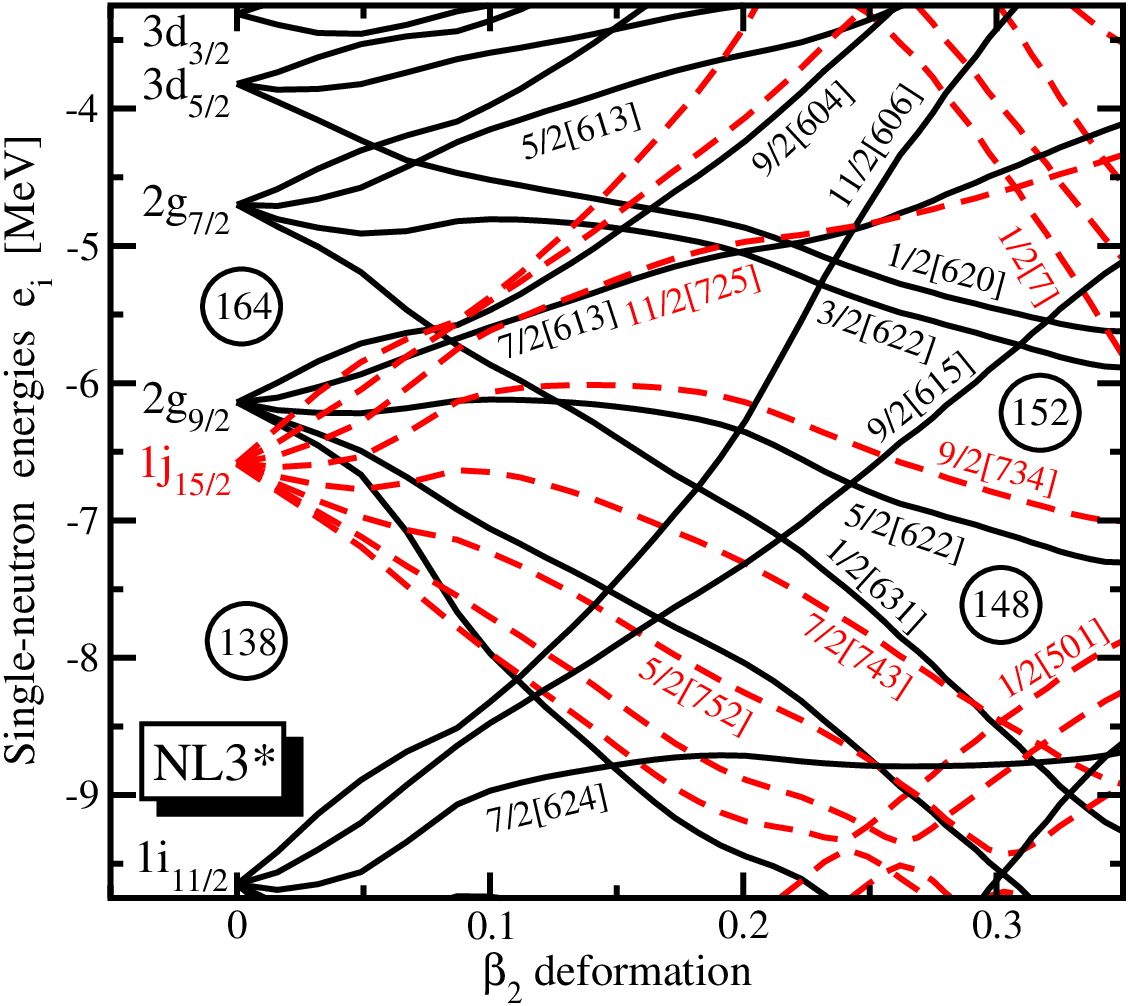}}
\caption{
\label{Nilsson-diagramsNL1}
Same as in Fig.~\protect\ref{no254.sly4.nilsson2} but for the covariant EDFs NL1 and NL3*.
All results and dominant Nilsson labels were determined using axially symmetric code \textsc{RHB}.
}
\end{figure}

The Nilsson diagrams, shown in
Figs.~\ref{no254.sly4.nilsson2}--\ref{Nilsson-diagramsNL1}, have been
obtained by diagonalizing the self-consistent mean-field Hamiltonians
corresponding to the states that were constraint to a sequence of values of the
axial mass quadrupole moment~\cite{[Rin80]},
$Q=\langle{2z^2-x^2-y^2}\rangle$. Then,
by using a simple phenomenological formula \cite{[Ram01]},
\begin{equation}\label{beta}
\beta_2 \equiv \frac{4\pi}{3R^2 A}\sqrt{\frac{5}{16\pi}}Q \simeq Q\times0.009\,\mbox{b}^{-1},
\end{equation}
values of the average quadrupole moments were translated into values of
the Bohr parameters $\beta_2$. The numerical factor of $0.009\,\mbox{b}^{-1}$
corresponds to $R=1.2$\,fm\,$A^{1/3}$ and $A=254$.
Note that the Nilsson diagrams obtained in this way constitute only
an illustration of single-particle properties of the nuclei in the nobelium
region.
\MiB{Indeed, going in either direction in the chart of nuclei, the relative
distances between spherical and deformed levels change, closing some of the
gaps visible in Figs.~\ref{no254.sly4.nilsson2}--\ref{Nilsson-diagramsNL1}
while opening others. Examples for such evolution of the shell structure
of spherical states and the deformed minima can be found in
Refs.~\cite{bender12,[Rut97],[Ben99a],[Shi14b]}. Small, but clearly visible
changes of the Nilsson diagrams can already be spotted when just adding a few
neutrons or protons, compare for example the Nilsson diagram obtained with
SLy4 for \nuc{254}{No}, Fig.~\ref{no254.sly4.nilsson2}, with the one for
\nuc{250}{Fm} presented in Refs~\cite{bender12}, or the Nilsson diagrams
obtained with NL1 and NL3* for \nuc{254}{No}, Fig.~\ref{Nilsson-diagramsNL1},
with those for \nuc{244}{Cm} presented in Ref.~\cite{[Afa13]}.
}

As we can see, for all considered EDFs, the overall positions and
de\-for\-mat\-ion-de\-pen\-dence of single-particle levels is fairly
similar. In particular, deformed shell gaps, which appear near
ground-state deformations of $\beta_2=0.3$,
occur at particle numbers of $Z=98$, $Z=104$ and $N=150$ and/or $N=152$
in the majority of the functionals. The only exception is the functional NL3*,
which is characterized by additional gaps at $Z=102$
and $N=148$. Another deformed gap at  $\beta_2=0.2$ is observed for $Z=110$. Moreover,
significant differences in important details are also visible. The
proton deformed shell gaps appear consistently above the one at
$Z=100$ that is tentatively inferred from the experimental data,
see discussion in Refs.~\cite{[Afa03],[Shi14a]}.

Interestingly, although different EDFs show similar deformed
neutron shell closures, we observe dramatic differences in the shell structure at
sphe\-ri\-ci\-ty between the Skyrme EDF UNEDF2 and the other ones, as shown in
Fig.~\ref{no254.sly4.nilsson2}. Compared to SLy4, spherical orbital
$1j_{15/2^-}$ ($1i_{11/2^+}$) is lowered (raised) by about 2\,MeV,
and thus their relative positions are inverted, resulting in
the spherical shell gaps at $N=152$ and 170, whereas other EDFs predict
shell gaps at $N=164$.

\MiB{The strong rearrangement of spherical neutron shells observed for UNEDF2
as compared to all other EDFs is a consequence of its rather large
$C^{JJ}_0$ and $C^{JJ}_1$ coupling constants of the spin-current tensor
terms of this parameterization \cite{[Kor14]}. In the $N \approx 152$ region,
the inversion of the spherical level sequence substantially increases the
number of filled spherical shells for which the spin-orbit partner is empty,
thereby increasing the size of the spin-current terms. In fact, such behaviour is
often found at mid-shell for parameterizations with large attractive tensor
terms \cite{[Les07],[Ben09b]}.}

\MiB{The relativistic NL1 and NL3* functionals have the unique feature that they
predict a large spherical $N=138$ gap of about 3\,MeV that is absent in
all non-relativistic calculations. As the sequence of spherical subshells
is different, for NL1 this gap is located between the $1i_{11/2^+}$ and
$2g_{9/2^+}$ levels, whereas for NL3* it is found between the $1i_{11/2^+}$
and $1j_{15/2^-}$ levels.}

\subsection{Quasiparticle spectra in $^{251}$Cf and $^{249}$Bk}
\label{sec3a}

\begin{figure}
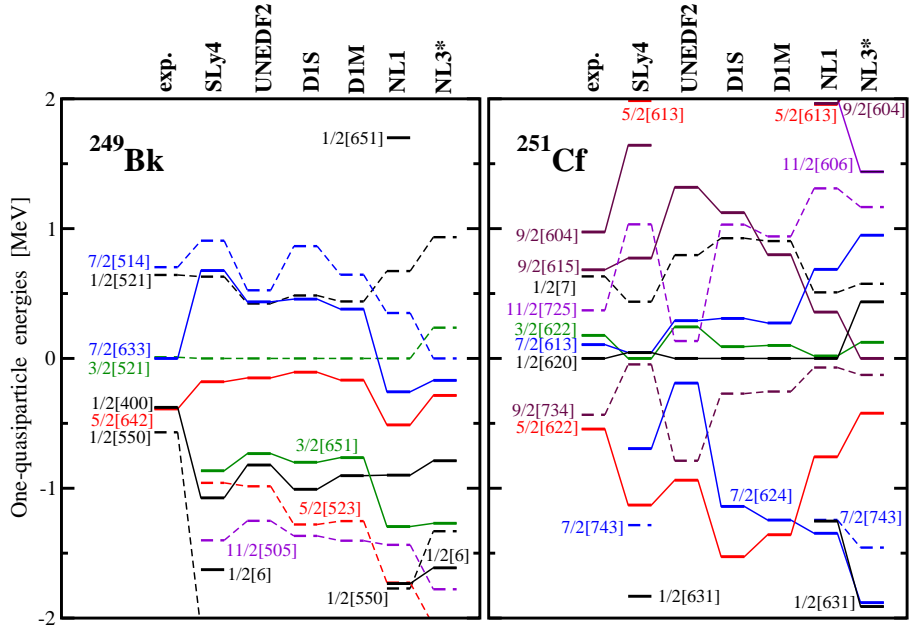
\hspace*{0.90cm}
\includegraphics[scale=0.4]{FigSHE04a.eps}\hspace*{-0.10cm}
\includegraphics[scale=0.4]{FigSHE04b.eps}
\caption[C1]{\label{cf151-bk29} Experimental and calculated  quasiparticle
spectra in $^{249}$Bk and $^{251}$Cf, see text for the convention used here.
Experimental data are taken from Ref.\ \protect\cite{Eval-data}
\AnA{We label the state with the full Nilsson label of the dominant
component of the wave function only if the squared amplitude of this
component exceeds 50\%. The exception is the 1/2[7] state which is
strongly mixed. However, the cumulative squared amplitude of
the components of the wave function with $N = 7$ in the structure of
this state exceeds 90\%. Thus, we label it only by principal quantum
number $N$ and $\Omega$.}}
\end{figure}

In Fig.~\ref{cf151-bk29}, calculated spectra of low-lying
band-heads in $^{249}$Bk and $^{251}$Cf are shown along with the
experimental data. In these results, the spectra were obtained
by individually blocking relevant quasiparticle orbitals
and then plotting differences of {\em total} many-body energies
of obtained nucleonic configurations with respect of the
total energy of the ground state, that is,
they are not at all equivalent to quasiparticle energies understood as
eigenvalues of the HFB Hamiltonian. The ground states were identified as
nucleonic configurations with the lowest total energies.

Nuclear configurations of deformed odd nuclei (one-quasiparticle
configurations) were labelled by means of the standard asymptotic quantum
numbers $\Omega [Nn_z \Lambda]$ (Nilsson quantum numbers) that correspond to the dominant
component in the wave function of the blocked quasiparticle state.

We used the convention of plotting positive (negative) values for the
excitation energies of quasiparticle configurations that correspond
to blocked quasiparticle states having norms of the second HFB
components smaller (larger) than 1/2 {\em before blocking}. In this
way, the states that are predominantly of a particle (hole) character
appear above (below) zero energy. Moreover, these states are always
plotted relatively to the ground-state; thus the ground state is
plotted identically at the value of zero energy. This convention
facilitates the comparison of fully self-consistent results with the
Nilsson diagrams. For experimental quasiparticle configurations, we
follow the assignments of particle/hole character as presented in
Figs.~34 and 35 of Ref.~\cite{[Jain90]}.

In these two odd nuclei, prominent intruder configurations correspond
to the proton 7/2[633] and neutron 11/2[725] orbitals. In
Ref.~\cite{[Shi14a]}, these two orbitals were used as benchmark
states to adjust strengths of the spin-orbit interactions. We see
that without such an adjustment, none of the studied standard EDFs
places them at the right position. The ground-state proton 7/2[633] orbital, which
experimentally is almost degenerate with the 3/2[521]
orbital,
for Skyrme and Gogny EDFs appears about 500\,keV above the
ground state and for covariant EDFs about 200\,keV below the ground
state, with the calculated ground states corresponding the 3/2[521]
orbital (or 7/2[514] for the covariant EDF NL3*).

The neutron 11/2[725] orbital, for covariant, Gogny, and SLy4 EDFs,
appears too high and for UNEDF2 EDF too low above its experimental
position with respect to the ground-state 1/2[620] orbital. On the
one hand, one can say that on the absolute scale these deficiencies
are not large. On the other hand, they may point to slightly
incorrect positions of spherical intruder orbitals, from which one
would like to infer the shell structure of as yet not-reached
superheavy nuclei.
This analysis shows that detailed structure of
very heavy deformed nuclei may depend on extremely fine details of
the present-day theoretical models, which very well may be far beyond
any reasonable possibility of adjusting them precisely enough to available
experimental data.

Similarly as in the analysis presented in Ref.~\cite{[Shi14a]}, as an
attempt to improve the agreement with the experimental values, we
have considered variations of the spin-orbit parameter $W_{LS}$ of
the Gogny EDF D1S that could influence relative positions of
intruder states. Increasing $W_{LS}$ from its nominal value of
130\,MeV\,fm$^{5}$ reduces the excitation energy of the $11/2^{-}$ state
while it increases the excitation energy of the $9/2^{-}$ state in
$^{251}$Cf. These changes improve the agreement with experimental
data for larger values of $W_{LS}$. In the $^{249}$Bk case, the
$7/2^{+}$ goes down in excitation energy as $W_{LS}$ increases, while
the $5/2^{+}$ and $1/2^{-}$ levels go up. As in the $^{251}$Cf case,
the comparison with experiment seems to favor larger values of
$W_{LS}$. This, however, has to be contrasted with the analysis
of the shell structure of heavy spherical nuclei like $^{208}$Pb, which
usually calls for weaker spin-orbit interaction \cite{[Ben03],[Ben99a],[Les07]}.

However, it is necessary to recognize that the studies
restricted to spin-orbit potential may have  internal limitations that
come from the fact that possible deficiencies in the description of
the energies of the single-particle states emerging from the central
potential, as for example those inferred in Ref.~\cite{chatillon06}, are ignored. The fact that standard Skyrme functionals
provide better description of the single-particle states in the
$Z=115$ nuclei than the ones with the strength of spin-orbit
interaction adjusted to experimental data in nobelium region
\cite{[Shi14b]} may be related to such limitation.

\begin{figure}
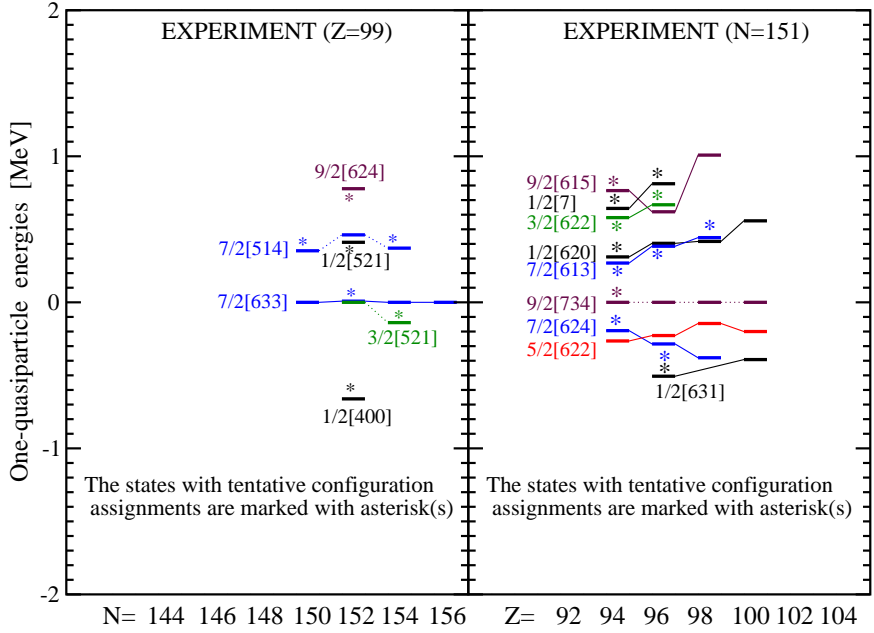
\hspace*{0.95cm}
\includegraphics[scale=0.45]{FigSHE05a.eps}\hspace*{-0.22cm}
\includegraphics[scale=0.45]{FigSHE05b.eps}
\caption[C1]{\label{EXP-151-99}
Experimental spectra \cite{Eval-data} of the even-$N$ $Z=99$
isotopes (left panel) and even-$Z$ $N=151$ isotones (right panel).
}
\end{figure}

To explore the sensitivity of the results to the amount of pairing
correlations in the system, for the Gogny EDFs we performed calculations where the
pairing strengths of protons and neutrons were multiplied by factors
$f_{p}$ and $f_{n}$, respectively. The first noticeable fact is
that increasing the neutron pairing strength does not influence in
a significant way the spectrum of $^{249}$Bk (odd $Z$) as it also happens
when increasing the proton pairing strength in $^{251}$Cf. Increasing
$f_{p}$ reduces the excitation energy of all levels except for the
$5/2^{+}$ state that remains more or less constant. The comparison
with experimental spectra seems to favor larger proton pairing
correlations. In the $^{251}$Cf case, all the levels except the
lowest $3/2^{+}$ decrease their excitation energy with increasing
pairing strength. As in the $^{249}$Bk case, increasing the pairing
correlations in $^{251}$Cf improves the agreement with experimental
spectra. The same behavior was observed during the readjustment of the pairing strength for SLy4. However, on should keep in mind that pairing strengths are
predominantly defined by the odd-even mass staggering, see
Sect.~\ref{sec3d} below.

\subsection{Quasiparticle spectra in $Z=99$ (Es) isotopes and $N=151$ isotones}
\label{sec3c}

\begin{figure}
\includegraphics[width=0.9\textwidth]{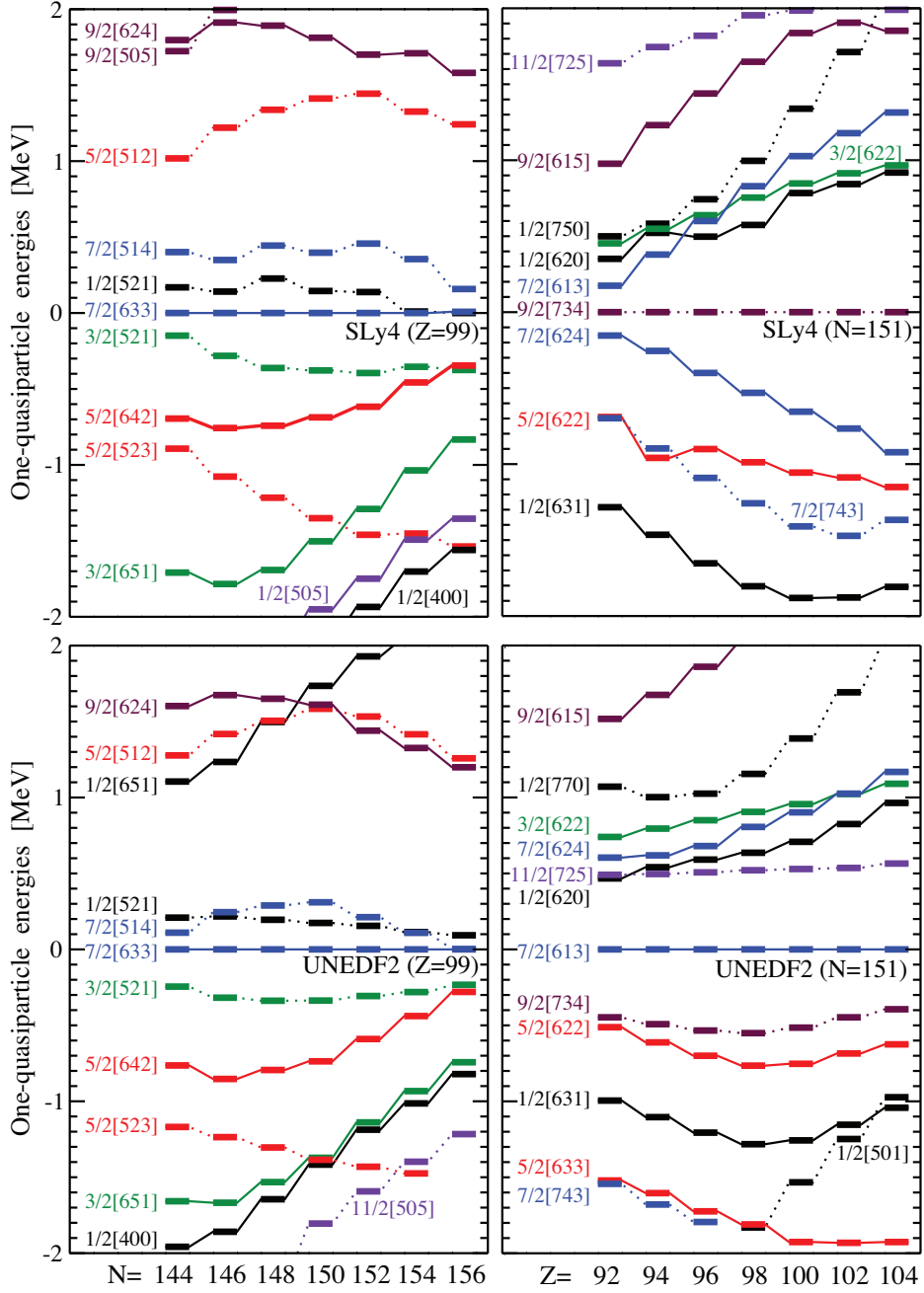}
\caption[C1]{\label{SLY4-151-99}
Same as in Fig.~\protect\ref{EXP-151-99} but for the spectra calculated
for the Skyrme EDF SLy4 (upper panels) and UNEDF2 (lower panels).
Since for SLy4 only the mean values of the single-particle
angular momenta $\langle \hat{j}_\parallel \rangle$ and
$\langle \hat{j}^2 \rangle$ (and not full
Nilsson labels) were available, in this case the labels
were assigned by analogy with the results obtained for the Gogny EDF D1S, which provides very similar spectra.}
\end{figure}

\begin{figure}
\includegraphics[width=0.9\textwidth]{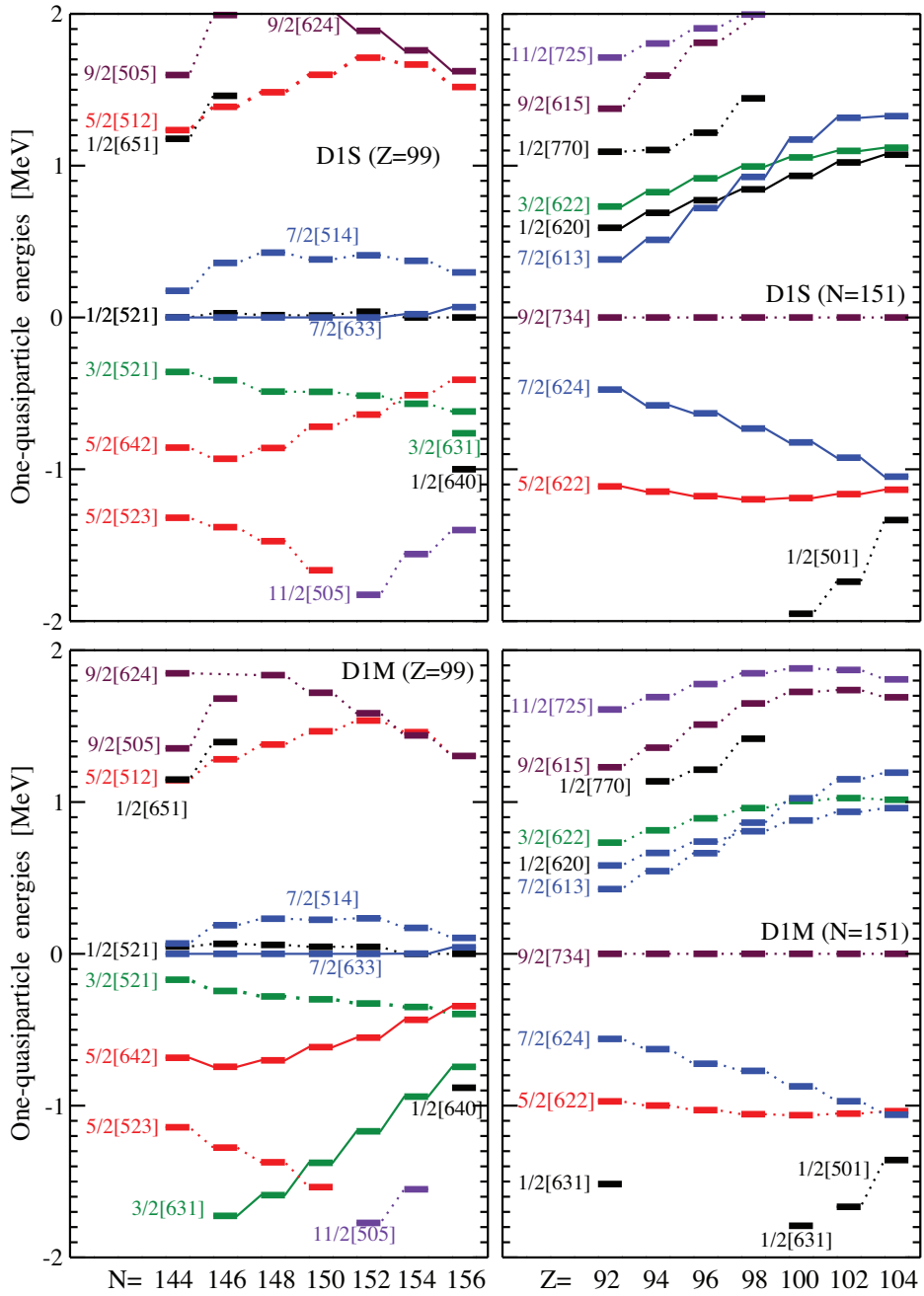}
\caption[C1]{\label{D1S-151-99}
Same as in Fig.~\protect\ref{EXP-151-99} but for the spectra calculated
for the Gogny EDF D1S (upper panels) and D1M (lower panels).
}
\end{figure}

\begin{figure}
\includegraphics[width=0.9\textwidth]{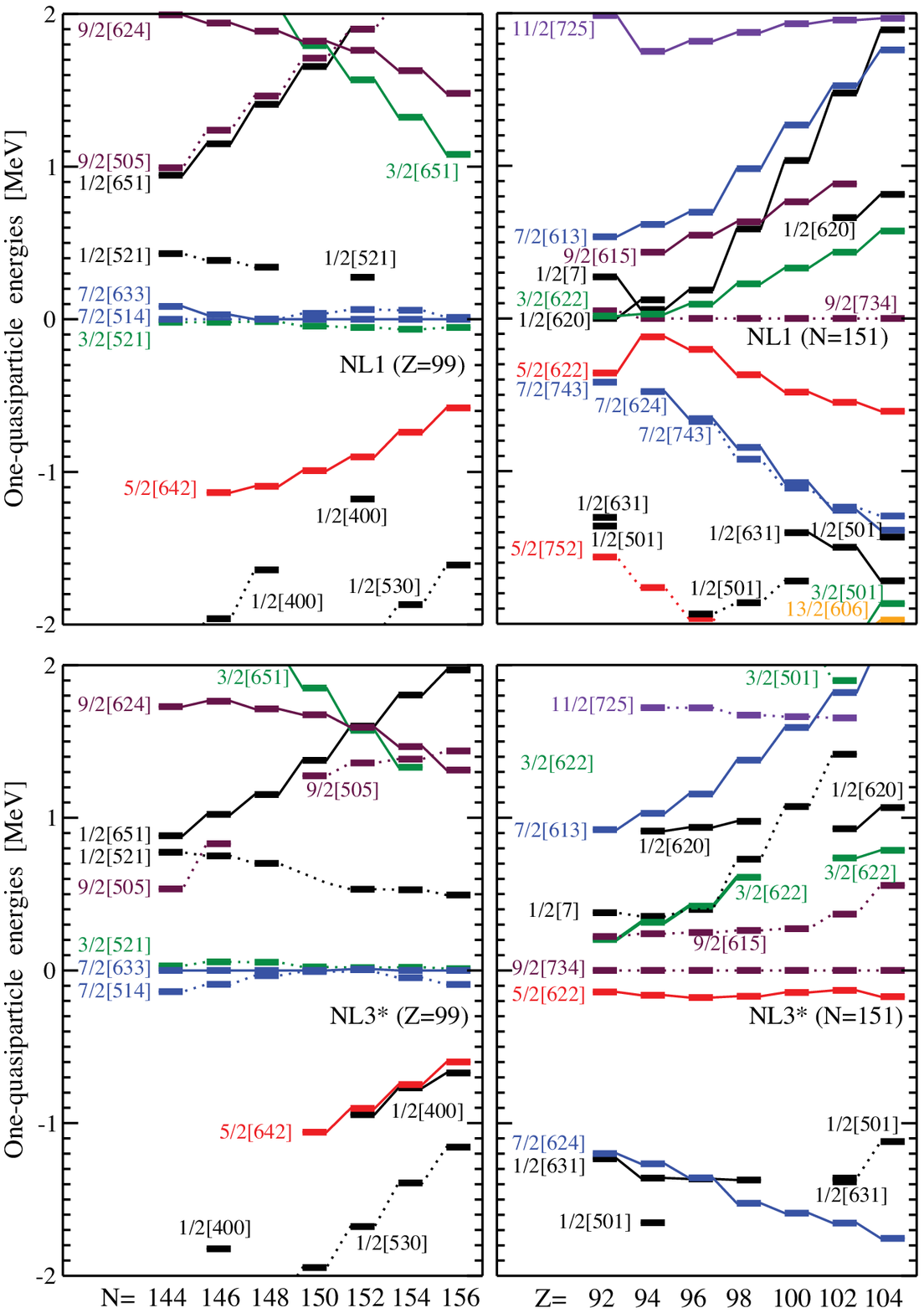}
\caption[C1]{\label{NL3-151-99}
Same as in Fig.~\protect\ref{EXP-151-99} but for the spectra calculated
for the covariant EDF NL1 (upper panels) and NL3* (lower panels).
}
\end{figure}

In the left (right) panels of
Figs.~\ref{EXP-151-99}--\ref{NL3-151-99}, we show experimental and calculated spectra of the even-$N$ $Z=99$
isotopes (even-$Z$ $N=151$ isotones). The results are presented in the
convention discussed in Sec.~\ref{sec3a}, with experimental assignments taken from Refs.~\cite{Eval-data,[Jain90]}. We see that the
experimentally assigned odd-quasiparticle configurations are
theoretically always obtained at relevant low excitation energies.
However, similarly as in the case of $^{249}$Bk and $^{251}$Cf
presented above, the details of the level spacing and ordering vary from
one calculation to another and are not very well reproduced.

Although in this region of nuclei, the experimental information is
richest in the particular isotopic and isotonic chains studied here,
it is still quite scarce, and often experimental assignments of
configurations are still only tentative \cite{Eval-data}.
Nevertheless, we can already
see several conspicuous experimental trends.

In protons, we see the 7/2[633] ground states and 7/2[514] excited
states at fairly constant excitation energies of about 400\,keV. For
the Skyrme EDF SLy4, this feature is very well reproduced, with a
significant drop of this excitation energy predicted at $N=156$. For
the Skyrme EDF UNEDF2, the 7/2[514] level is also obtained above
the Fermi level, with the excitation energy gradually decreasing
already at $N=152$. Note, however, that for the Skyrme EDFs, in
$^{249}$Bk the relative positions of these two levels were not very
precisely reproduced, so the nice agreement obtained in the $Z=99$
isotopes might be fortuitous. For the Gogny EDFs, this pair of the levels
is obtained at roughly correct excitation energies,
whereas for the
covariant EDFs, these levels are fairly well degenerate.

In neutrons, the 9/2[734] ground states are for all EDFs studied here
well reproduced, apart from the Skyrme EDF UNEDF2, which gives the
7/2[613] ground state with the hole-character 9/2[734] orbitals
appearing at excitation energy of about 400-500\,keV. Two
particle-character excited quasiparticle states,  7/2[613] and 1/2[620],
show excitation energies which increase with mass. They are correctly
reproduced for the Skyrme SLy4 and covariant EDF's; however, the
increase with mass is too fast.
Correct trends of these two
levels are also determined for the Gogny EDFs. Two other hole-character
experimental quasiparticle states, 7/2[624] and 5/2[622], have
excitation energies weakly increasing and decreasing with mass,
respectively. For the Skyrme SLy4 and covariant NL1 EDFs, their
excitations energies both increase quite rapidly with mass, whereas for
the Gogny functionals the mass dependence is rather correct. In all models
studied here, the energy splitting of these two levels is too large
as compared to data.

\MiB{From the systematics of the one-quasiparticle levels one can draw a
number of interesting conclusions about the significance of the
Nilsson diagram of single-particle levels for the calculated and
observed spectra of one-quasi\-par\-ticle levels. Already in schematic
models there is no quantitative one-to-one correspondence, as the presence of
pairing correlations modifies the spectrum of low-lying states.
In nuclear EDF calculations, there are additional self-consistency
effects from the separate optimization of each one-quasiparticle state.
}

\MiB{As already outlined in Sect.~\ref{sec3b}, the various EDFs do not always
agree on the size of the energy gaps between spherical subshells in the
Nilsson diagrams for \nuc{254}{No} plotted in Figs.~\ref{no254.sly4.nilsson2},
\ref{no254.d1s.nilsson2} and~\ref{Nilsson-diagramsNL1}, sometimes even
not on the sequence of the levels.
For example, the relativistic NL1 and NL3* functionals predict a large
spherical $N=138$ gap of about 3\,MeV. For the non-relativistic D1S, D1M,
and SLy4 EDFs this gap is much smaller and even disappears
for UNEDF2. With the non-relativistic functionals one finds spherical
gaps at $N=148$ and $N=152$ instead. The appearance and size of the
$N=138$ gap is controlled by the position of the spherical $1i_{11/2^+}$
level below and the $2g_{9/2^+}$ level above, except for NL3* for which
the $1j_{15/2^-}$ is pulled below the $2g_{9/2^+}$.
The neutron 9/2[734], 5/2[622], and 9/2[615] single-particle levels that
emerge from the spherical $1j_{15/2-}$, $2g_{9/2^+}$, and $i_{11/2^+}$
subshells, respectively, are found just above and below the deformed
$N=150$ and $N=152$ gaps in the Nilsson diagrams for \nuc{254}{No}.
Looking at one-quasiparticle spectra of \nuc{251}{Cf},
as plotted in Fig.~\ref{cf151-bk29},
one finds that for SLy4, D1S, and D1M the distance between the
hole-character 9/2[734] and particle-character 9/2[615] levels is very
satisfactorily described within 200\,keV, whereas for UNEDF2 it is much
too large, and for NL1 and NL3* it is much too small.
On the other hand, in the $N=151$ isotones shown in Figs.~\ref{EXP-151-99}--\ref{NL3-151-99}, the distance
between these two levels is on average too large for SLy4, D1S, and D1M, again much
too large for UNEDF2, correct for NL1, and again much too small for NL3*.
This indicates that the spherical $N=138$ gap of NL3*
is probably too large. Still, it is unlikely that it
has to be made as small as the one found for Skyrme and Gogny EDFs.
}

\MiB{By contrast, the distance between the 9/2[734] and 5/2[622] one-quasi\-particle
levels in \nuc{251}{Cf} and $N=151$ isotones, which are connected to the $1j_{15/2^-}$ and $2g_{9/2^+}$ shells,
respectively, is
overestimated by almost 1\,MeV for SLy4, D1S, and D1M, by a few hundred keV for NL1 and UNEDF2, but quite well
reproduced by NL3*. This indicates first
of all that the pronounced deformed $N=150$ gap visible in the Nilsson
diagrams for SLy4, D1S, D1M, and NL1 should be much smaller. In fact, the
large deformed gap at $N=150$ that is predicted by a substantial number of
nuclear EDFs has already quite often been attributed to be one of the main
causes for the disagreement between calculated and observed spectroscopic
properties of nuclei in the $A \approx 250$ mass region
\cite{[Her08],[Afa03],[Shi14a],{chatillon06},bender12}.
It is remarkable that the UNEDF2 and NL3* functionals that both
describe well the distance between the 9/2[734] and 5/2[622] levels give
very different shell structure at spherical shape in the Nilsson diagrams
of Figs.~\ref{no254.sly4.nilsson2} and~\ref{Nilsson-diagramsNL1}. For UNEDF2,
the spherical $1j_{15/2-}$ shell is about 500\,keV above the spherical
$2g_{9/2^+}$, whereas for NL3* it is the other way round, the latter being
a unique feature among the functionals studied here. Note that for NL3*
the relative position of these levels is
quickly moving with particle number: for the slightly lighter \nuc{244}{Cm}
the $1j_{15/2-}$ is already above $2g_{9/2^+}$ level like for the other
functionals studied here, cf.\ Ref.~\cite{[Afa13]}.

Altogether, these findings indicate that on cannot expect to find a unique
one-to-one correspondence between the spectra of one-quasiparticle levels of
deformed nuclei and spherical single-particle levels in a Nilsson diagram.
Different shell structure in the Nilsson diagrams might lead to similar
one-quasiparticle spectra and vice versa. Indeed, the rearrangement,
polarization, pairing and single-particle-mixing effects when constructing
self-consistent one-quasiparticles state make the connection quite complex
and apparently also slightly EDF-dependent.
In addition, it should be recalled that for nuclei in the $A \approx 250$
region the spherical configuration corresponds to a maximum of the deformation
energy landscape and therefore should not be associated with a physical
state.
}

\MiB{Similar cases can be found for proton levels. All Nilsson diagrams for \nuc{254}{No}
plotted in Figs.~\ref{no254.sly4.nilsson2}, \ref{no254.d1s.nilsson2}
and~\ref{Nilsson-diagramsNL1} exhibit a substantial spherical $Z=92$ gap that
is located between the $2f_{7/2^-}$ and $1h_{9/2^-}$ orbitals in non-relativistic functionals
and between the $1i_{13/2}$ and $1h_{9/2}$ orbitals in covariant functionals.
It has been
pointed out in Refs.~\cite{216Th,218U} that there is no indication for such
a gap, which is also visible in the single-particle spectra of lighter spherical
nuclei \cite{[Ben99a]}, in the available spectroscopic data for the spherical
\nuc{216}{Th} and \nuc{218}{U} nuclei.
However, the $7/2[514]$, $3/2[521]$, $1/2[521]$, and $7/2[633]$
one-quasiparticle levels of \nuc{249}{Bk} and the Es isotopes, which
originate from the spherical $1h_{9/2}$, $2f_{7/2}$ and $1i_{13/2}$
subshells surrounding the spherical $Z=92$ gap,
are reasonably well described by all EDFs employed here, see Figs.~\ref{cf151-bk29}--\ref{NL3-151-99}.
As a consequence, a large spherical $Z=92$ gap in the Nilsson diagram of
\nuc{254}{No} is not in apparent conflict with the available data for
deformed nuclei in the $A \approx 250$ mass region. In particular the
non-relativistic Skyrme and Gogny functionals describe well the
relative position of the $3/2[512]$ and $7/2[514]$ levels within a few hundred keV. It is
only for the relativistic NL1 and NL3* functionals that the spacing
between the $7/2[514]$ and $7/2[633]$ one-quasiparticle levels
becomes slightly too small, cf.\ Figs.~\ref{cf151-bk29} and~\ref{NL3-151-99},
which points to a slight overestimation of the spherical $Z=92$ gap
for these functionals. Still, the necessary shift of the spherical shells
that one would expect to correct for the disagreement between calculation and
data will not even reduce the gap to the size found with non-relativistic
EDFs.
}

\AnA{It is necessary to recognize that the present investigation represents
one of first steps in the direction of understanding of the accuracy of the
description of one-quasiparticle states in deformed nuclei, which,
however, goes beyond previous attempts by directly comparing different
classes of the  EDF approaches for the same set of experimental data. Based on the
current set of experimental data similar accuracy is achieved for employed
EDF's. So far statistical analysis of the accuracy of the description of
one-quasiparticle states employing full set of available experimental data
on proton and neutron states has been performed only in actinides and only
in the framework of CDFT using NL1 and NL3* EDF's in Ref.\ \cite{AS.11}.
Although many of the states are described with acceptable accuracy, for
some states the deviation of calculated energy from experimental one
exceeds 1 MeV.}

\AnA{Less systematic studies have been performed in the non-relativistic
EDF approaches. Global survey of the ground state configurations in odd-mass
nuclei employing three Skyrme functionals has been performed in Ref.\
\cite{BQM.07}. In Skyrme EDF, the spectra of few actinides and of odd-proton
Ho nuclei have been studied in Refs.\ \cite{[Ben03d],[Sch10]}. The spectra
of selected Rb, Y, and Nb nuclei have been studied in axial Gogny EDF
in Refs.\ \cite{RSR.10,RSR.11}.}

\AnA{These investigations reveal the same two sources of uncertainties
\cite{AS.11}. Although the same set of single-particle states appear
in the vicinity of the Fermi level as in experiment, their relative
positions and energies are not always correct. This is first source
of uncertainties. The second source of uncertainty is related to
stretched energy scale in model calculations as compared with
experiment which is due to low effective mass of the nucleon at
the Fermi level.}

\AnA{While the solution of the first problem can be attempted in the EDF
framework, the current analysis with different classes of EDF approaches suggest
that it is not likely to remove all existing problems in the description
of the single-particle spectra (see also the discussion in Ref.\
\cite{AS.11}).  More comprehensive solution, which would
also address the second source of uncertainty,  would
require taking into account particle-vibration coupling which will
explain existing fragmentation of the single-particle states and possibly
compress the calculated one-quasiparticle spectra bringing them closer
to experiment. Combined with respective re-parametrization of the
functionals it may lead to the functionals with better spectroscopic
quality. It is clear that existing functionals are biased towards
bulk properties since either no (CDFT) or extremely limited information
on single-particle properties (Skyrme and Gogny EDF) is used in their
fitting protocols. Even in this situation the calculated spectra are
not far away from experiment which can be considered as a success and
a good starting point for future development.}

\subsection{Odd-even and two-particle mass staggering}
\label{sec3d}

\begin{figure}
\centerline{\includegraphics[width=0.65\textwidth,angle=270]{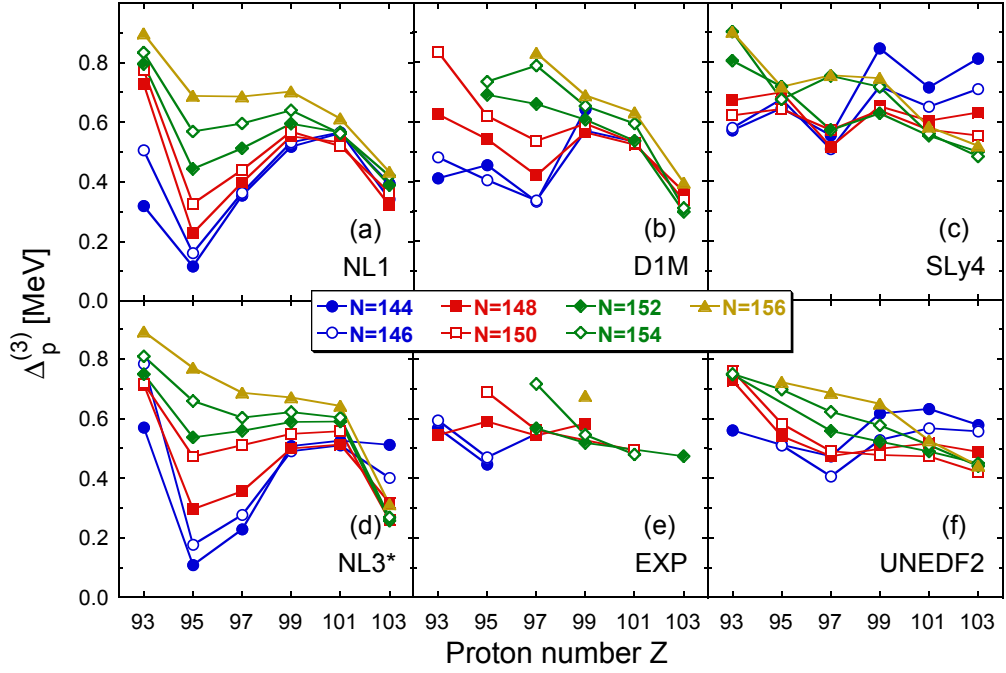}}
\caption[C1]{\label{DP}
Three-point proton odd-even mass staggering, Eq.~(\protect\ref{D3P}),
shown for the odd-$Z$ and even-$N$ nuclei in the nobelium region.
Experimental values are based on the AME2012 atomic mass evaluation
\cite{[Aud12]}.
}
\end{figure}

\begin{figure}
\centerline{\includegraphics[width=0.65\textwidth,angle=270]{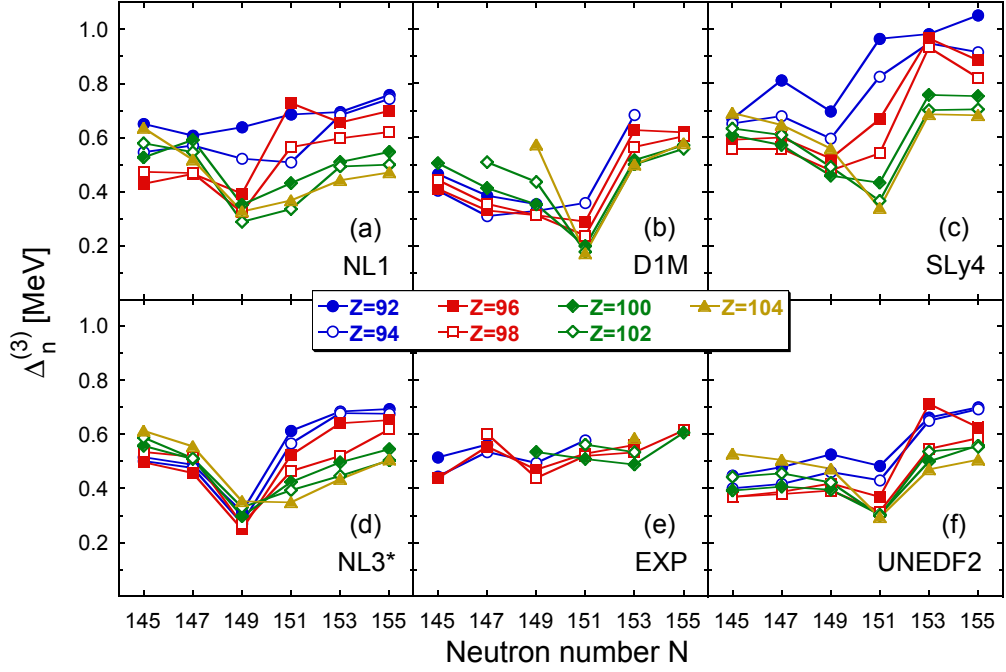}}
\caption[C1]{\label{DN}
Same as in Fig.~\protect\ref{DP} but for the neutron odd-even mass staggering, Eq.~(\protect\ref{D3N}),
shown for the odd-$N$ and even-$Z$ nuclei.
}
\end{figure}

In order to analyze the odd-even mass differences, the three-point
pairing indicators~\cite{[Sat98a]} (staggering parameters),
\begin{equation}\label{D3P}
\Delta_{p}^{(3)}=\frac{1}{2}\Big(B(Z+1,N)+B(Z-1,N)-2B(Z,N)\Big)
\end{equation}
with odd $Z$ and even $N$, and
\begin{equation}\label{D3N}
\Delta_{n}^{(3)}=\frac{1}{2}\Big(B(Z,N+1)+B(Z,N-1)-2B(Z,N)\Big)
\end{equation}
with even $Z$ and odd $N$, where $B(Z,N)$ is the (positive) binding energy of the nucleus,
have been plotted in Figs.~\ref{DP}
and~\ref{DN}, respectively.
In our HFB and RHB calculations, increased sizes of the shell gaps can, in principle,
be seen in the odd-even mass staggering parameters (\ref{D3P}) and
(\ref{D3N}), because the density of single-particle states has an
immediate bearing on the calculated intensity of pairing
correlations. This is under assumption that the differences obtained by
blocking different orbitals along the isotopic or isotonic chains have
a lesser impact on the odd-even mass staggering. Irrespective of the
detailed reproduction of the experimental values of these parameters,
calculated results may thus
illustrate the shell structure corresponding to the EDFs
studied here.

Experimental results, shown in panels (e) of Figs.~\ref{DP}
and~\ref{DN}, indicate
that in the nobelium region, values of the staggering parameters are
within the range of 500--700\,keV. On closer inspection, we see
several trends in the mass dependence of these parameters, which may
indicate variations of the shell structure due to the influence of
the level density on pairing correlations, or due to other fine
structural effects. In particular, values of $\Delta_{p}^{(3)}$ seem
to have a small dip for the $N=146$ isotones at $Z=95$, are fairly
constant in the $N=148$ isotones, and gradually decrease with mass in
the $N=150$--154 isotones. None of these values indicate a particularly significant
shell-gap opening near $Z=100$. Similarly, small dips in
$\Delta_{n}^{(3)}$, which show up in the $Z=96$--98 isotopes at
$N=149$ and in the $Z=100$ isotopes at $N=153$, do not point to a
particularly large shell gap at $N=152$.

\begin{figure}[tb]
\centerline{\includegraphics[width=0.65\textwidth,angle=270]{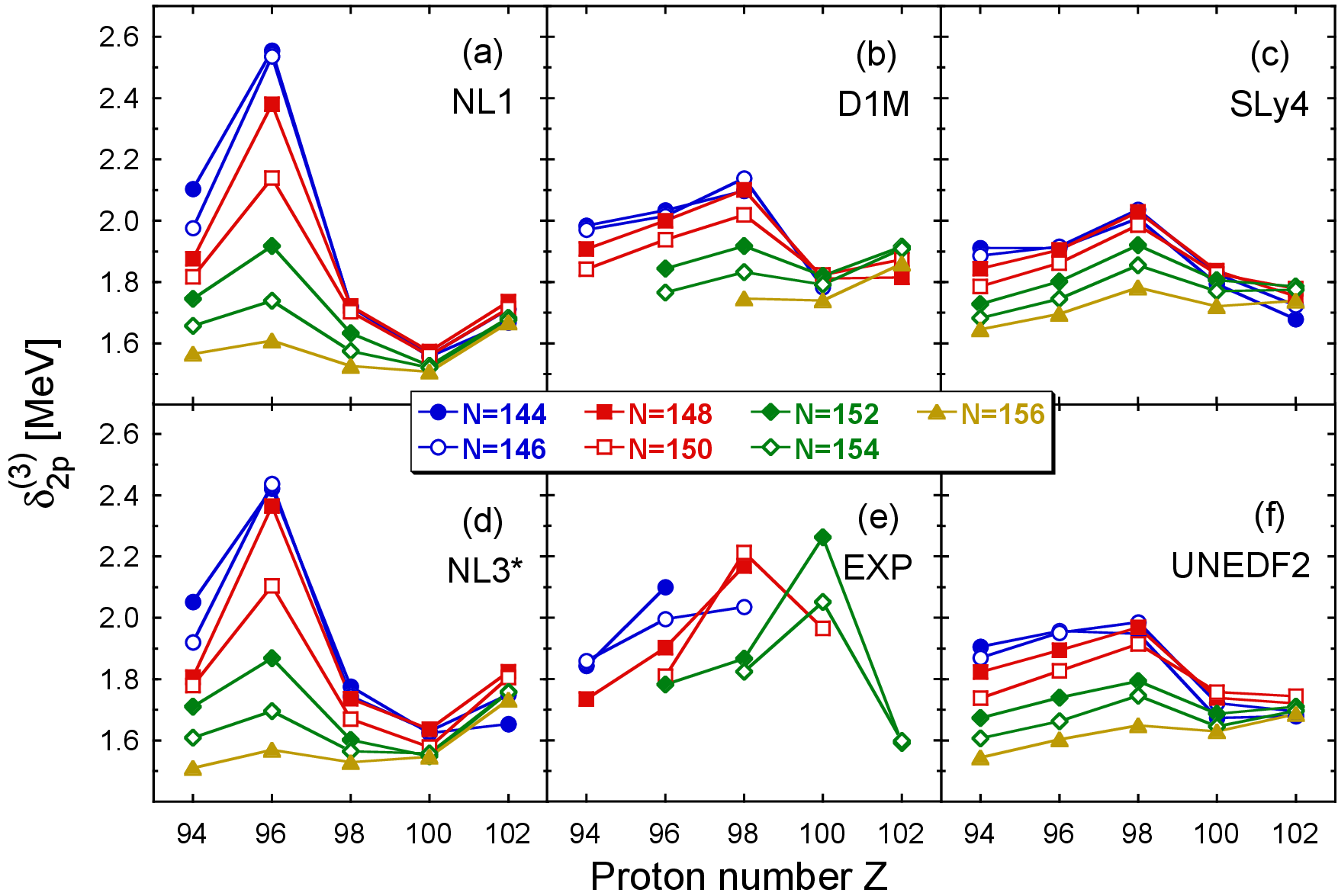}}
\caption[C1]{\label{D2P}
Two-proton
mass staggering, Eq.~(\protect\ref{D3P2}),
shown for the even-$Z$ and even-$N$ nuclei in the nobelium region.
Experimental values are based on the AME2012 atomic mass evaluation
\cite{[Aud12]}.
}
\end{figure}

\begin{figure}[tb]
\centerline{\includegraphics[width=0.65\textwidth,angle=270]{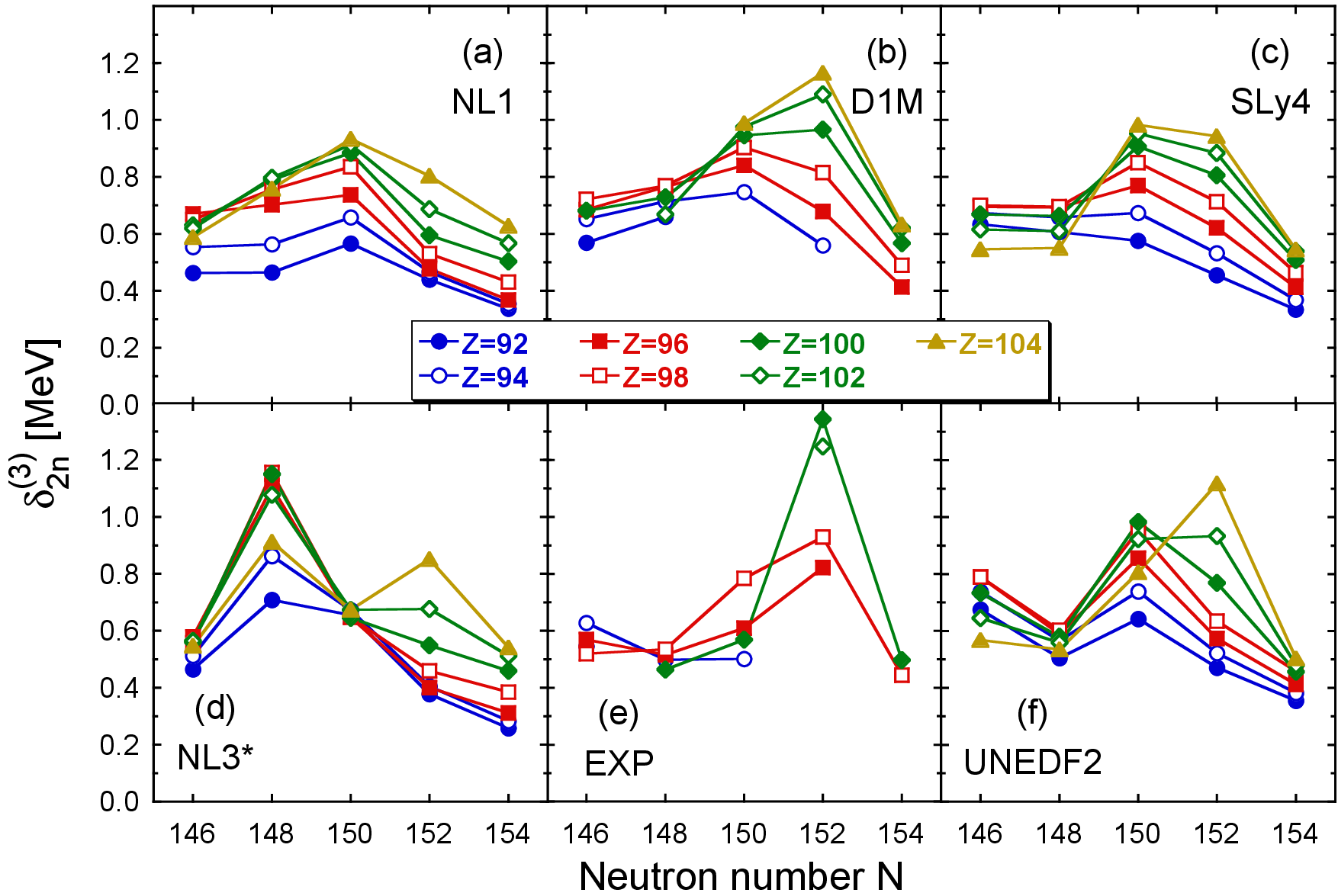}}
\caption[C1]{\label{D2N}
Same as in Fig.~\protect\ref{D2P} but for the two-neutron
mass staggering, Eq.~(\protect\ref{D3N2}).
}
\end{figure}

This lack of large
variations in odd-even mass staggering is at variance with the analysis of two-particle
mass staggering given by quantities
\begin{equation}\label{D3P2}
\delta_{2p}^{(3)}=2B(Z,N)-B(Z+2,N)-B(Z-2,N) = S_{2p}(Z,N)-S_{2p}(Z+2,N)
\end{equation}
and
\begin{equation}\label{D3N2}
\delta_{2n}^{(3)}=2B(Z,N)-B(Z,N+2)-B(Z,N-2) = S_{2n}(Z,N)-S_{2n}(Z,N+2) ,
\end{equation}
which were typically used to identify two-nucleon shell gaps in experiment and in calculations
for spherical \cite{[Rut97],[Li14]} and deformed \cite{[Bur98],[Afa03]} shell
closures in the predictions of Skyrme and covariant EDFs. As discussed in
Ref.~\cite{[The15]}, experimental values of $\delta_{2n}^{(3)}$ show clear maxima
for the $Z=96$--102 isotopes at $N=152$ and those of
$\delta_{2p}^{(3)}$ exhibit maxima for the $N=148$--150 isotones at
$Z=98$ and for the $N=152$--154 isotones at $Z=100$, see also Figs.~\ref{D2P} and~\ref{D2N}.

When looking at the most pronounced features of the calculated odd-even mass staggering
shown in Figs.~\ref{DP} and~\ref{DN}, we see that minima of
$\Delta_{p}^{(3)}$ can be seen at $Z=95$ (NL1 and NL3* EDFs) and $Z=97$
(D1M, SLy4, and UNEDF2 EDFs). For the Gogny and Skyrme
EDFs, these minima disappear at higher neutron numbers and rather
monotonic trends are then obtained. Similarly, minima of
$\Delta_{n}^{(3)}$ appear at $N=149$ (NL1 and NL3* EDFs) or $N=151$
(D1M, UNEDF2, and SLy4 EDFs); in the latter case, in lighter isotopes they
tend to shift to $N=149$. For the calculated two-proton-staggering indicators (\ref{D3P2}),
covariant EDFs, NL1 and NL3*, exhibit very strong maxima at $Z=96$, at variance with the data,
whereas the non-relativitic EDFs, D1M, SLy4, and UNEDF2, reproduce experimental maxima at $Z=98$
in the $N=146$--150 isotones but fail to shift these maxima to $Z=100$ in heavier isotones.
This conspicuous experimental fearure thus remains unsolved.
The calculated two-neutron-staggering indicators (\ref{D3N2}),
do not reproduce experimental maxima occurring at $N=152$.
These results illustrate the fact
that none of the studied EDFs reproduces the experimental trends in
shell gaps extracted from the two-particle indicators (\ref{D3P2})
and (\ref{D3N2}). We note here that the inclusion of the LN method into the calculations
renders pairing correlations much less sensitive to the shell structure.
Therefore, one then obtains fairly structureless trends of
$\Delta_{p}^{(3)}$ and $\Delta_{n}^{(3)}$ \cite{[Shi14a]}, although
for covariant EDFs, one at the same time obtains a significant
improvement of the overall agreement with experimental values \cite{[Afa13]}.

\subsection{Moments of inertia}
\label{sec3e}

The moments of inertia may constitute another independent indicator
of the shell structure. This is true under the assumption that
their values, similarly as for the odd-even staggering,
are dictated by varying pairing correlations. Then, larger shell
gaps would induce weaker pairing and thus larger moments of inertia.

The kinematical moments of inertia were calculated as
\begin{equation}\label{J1th}
J^{(1)} \equiv \frac{\langle \hat{J}_\perp \rangle}
            {\omega_\perp} ,
\end{equation}
where $\omega_\perp$ is the value of the constrained rotational frequency
and $\langle \hat{J}_\perp \rangle$ is the expectation value of the component
of angular momentum, both in the directions perpendicular to the axial-symmetry axis.
The dynamic moment of inertia is not considered due to larger experimental
uncertainties consequence of higher order derivatives in its definition.
Moreover, contrary to the kinematic moment of inertia the dynamic one
does not depend on spin. Thus, the kinematic moment of inertia provides
stricter constrain of model predictions.

\begin{figure}[tb]
\includegraphics[width=\textwidth]{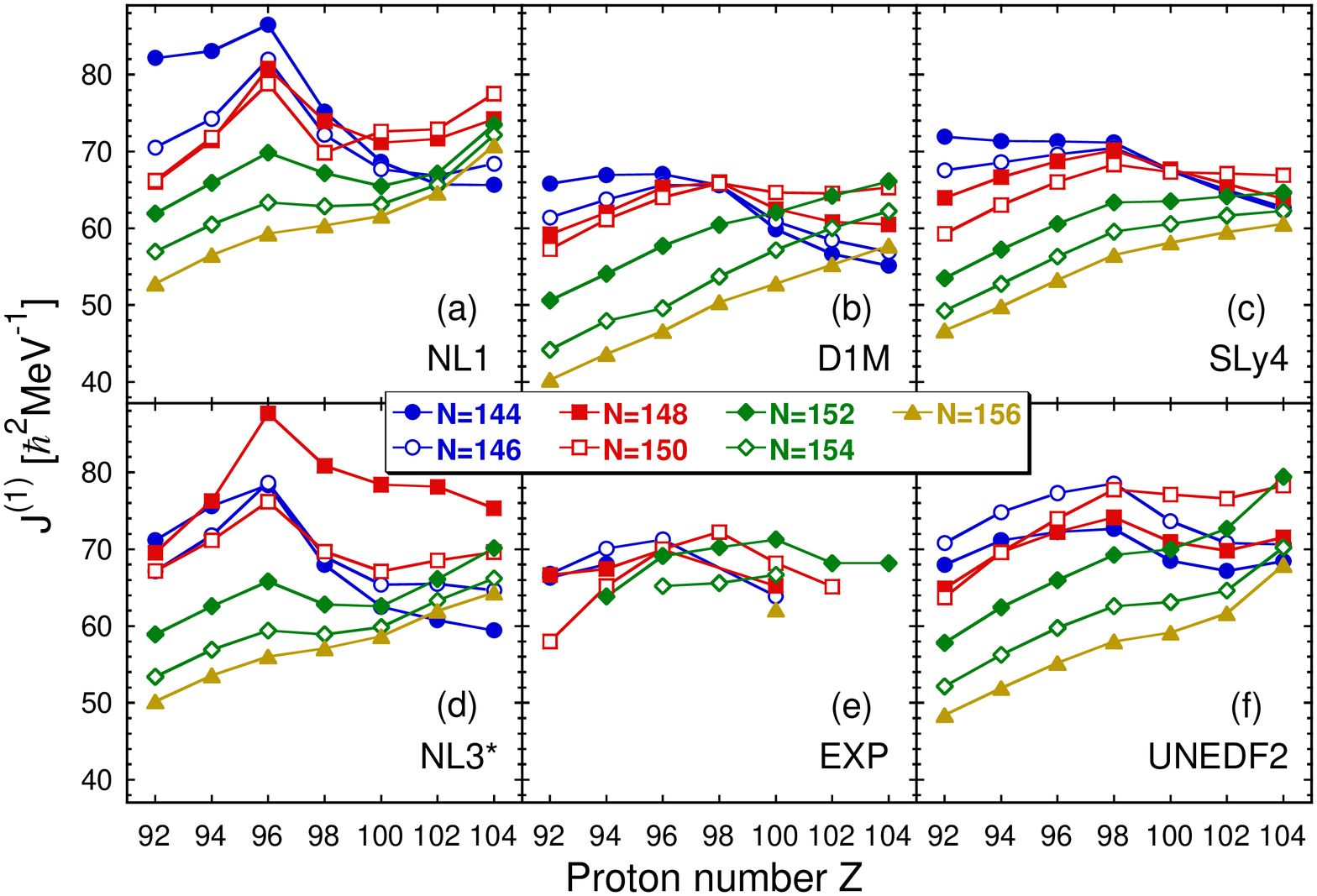}
\caption[C1]{\label{J1P}
Kinematical moments of inertia $\mathcal{J}^{(1)}$ of yrast rotational
bands of even-even nuclei in the nobelium region plotted in the isotonic chains
as functions of $Z$. Theoretical values, calculated at the
angular frequency of $\hbar\omega_\perp=20$\,keV, Eq.~(\protect\ref{J1th}),
are compared to experimental values
extracted from energies of the 2$^+$ states, Eq.~(\protect\ref{J1ex}).
Experimental energies are taken from Ref.~\cite{Eval-data} except
for $^{252}$Fm, which is taken from Ref.~\cite{[Asa15]}.
}
\end{figure}

\begin{figure}[tb]
\includegraphics[width=\textwidth]{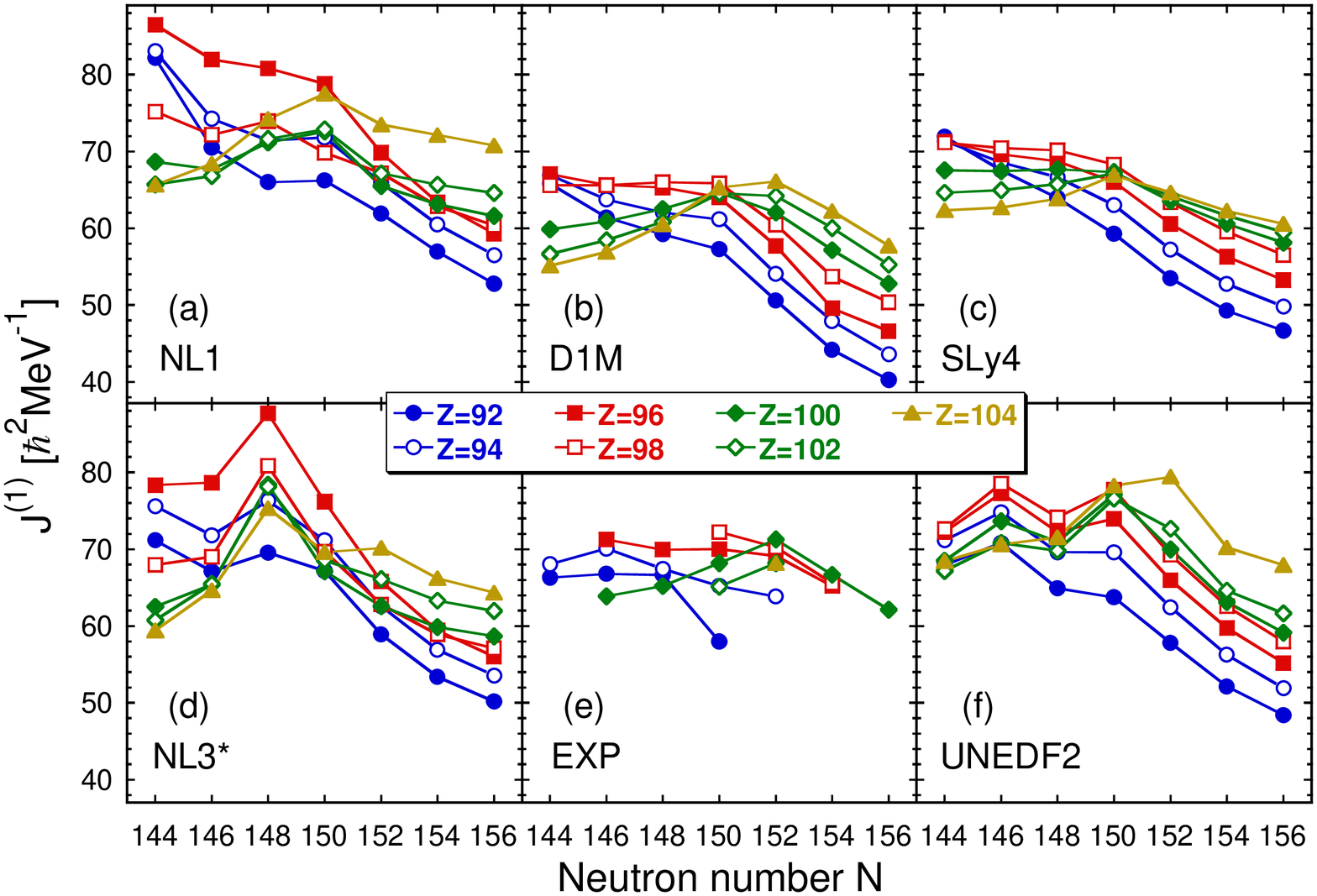}
\caption[C1]{\label{J1N}
Same as in Fig.~\protect\ref{J1P}, but plotted along the isotopic chains as functions of $N$.}
\end{figure}

All calculations of moments of inertia were performed at $\hbar\omega_\perp=20$\,keV.
In Figs.~\ref{J1P} and \ref{J1N}, they are compared to experimental values
determined as
\begin{equation}\label{J1ex}
J^{(1)} \equiv \frac{3\hbar^2}{E_{2^+}} .
\end{equation}
In nuclei where at the bottom of rotational bands the experimental
$2^+$ levels have not been seen, we used values
extrapolated~\cite{[Gre12]} by the Harris formula. In this region of
nuclei, the low-spin moments of inertia turn out to be very weakly
dependent on the angular frequency, and, therefore, a specific method
of extracting them from experiment is not essential.

The experimental values shown in panels (e) of Figs.~\ref{J1P} and
\ref{J1N} show clear maxima of $J^{(1)}$ in function of $Z$ at
$Z=100$ in the $N=152$ isotonic chain, and at $Z=98$ in the $N=150$
isotonic chain, as well as function of $N$ at $N=152$ in the
$Z=100$ isotopic chain. These maxima only partly appear in those
nuclei that show maxima of the two-particle staggering indicators,
discussed in Sec.~\ref{sec3d}.

In our theoretical calculations, weak maxima of $J^{(1)}$ are
obtained at $Z=98$ for $N=146$--150 isotonic chains (D1M\footnote{Results for D1S can be found in Ref.~\cite{[Del06]}}, UNEDF2, and
SLy4 EDFs) and a stronger maxima at $Z=96$ for $N=146$--152 isotonic
chains (NL1 and NL3* EDFs), whereas in heavier isotonic chains we see only
a gradual increase, without indications of increased shell gaps.
Similarly, in the isotopic chains maxima appear at $N=148$ (NL1 EDF)
and merely kinks appear at $N=150$ (D1M, UNEDF2, and SLy4 EDFs)

Comparing moments of inertia, Figs.~\ref{J1P} and \ref{J1N}, with the Nilsson diagrams,
Figs.~\ref{no254.sly4.nilsson2}--\ref{Nilsson-diagramsNL1}, we see that our calculations
with different models and forces seem to exhibit rather nice correspondence
between the respective proton (neutron) single-particle shell gaps and peaks/kinks
in the moments of inertia along the isotonic (isotopic) chains.

In Fig.~\ref{J1P}, for the covariant EDFs NL1 (a) and NL3* (d), the peaks at $Z=96$ obtained for the
$N=144$--152 isotonic chains can be associated with the shell gap that in Fig.~\ref{Nilsson-diagramsNL1} opens up at
$Z=96$. The peak moves to $Z=104$ for $N=154$ and 156 chains, see
Fig.~\ref{Nilsson-diagramsNL1}. For the Skyrme EDFs SLy4 (c) and UNEDF2 (f), the peak/kink at
$Z=98$ may be associated with a shell gap at $Z=98$ visible in the Nilsson diagram of
Fig.~\ref{no254.sly4.nilsson2}. For the Gogny EDF D1M, the correspondence is not as
clear as that visible in other cases. It is to be noted that all our calculations
predict that for most of the isotonic chains, values of $\mathcal{J}^{(1)}$ peak at $Z=104$.
This corresponds to the proton gap at $Z=104$ that is clearly visible in
Figs.~\ref{no254.sly4.nilsson2}--\ref{Nilsson-diagramsNL1}.

In Fig.~\ref{J1N}, for the covariant EDF NL3* (d), values of $\mathcal{J}^{(1)}$ show pronounced maxima at $N=148$ for isotopic
chains of $Z=96$--104. For NL1 (a), the maximum becomes a kink occurring at $N=150$.
Neutron numbers of $N=148$ and 150 correspond nicely to the shell gaps shown in
Fig.~\ref{Nilsson-diagramsNL1}.
Non-relativistic functionals predict either a peak at $N=150$ or a plateau for
$N=150$ and 152, particularly for the $Z=100$--104 isotopic chains, shown in Fig.~\ref{J1N}.

It is necessary to recognize that for a proper reproduction of
experimental moments of inertia, the inclusion of the LN method into
the calculations appears to be more important in covariant \cite{[Afa00c]} than
in non-relativistic EDFs. The LN method renders the values of calculated
moments of inertia much closer to the data \cite{[Afa13],A.14} and at the same
time much less sensitive to the underlying shell structure, at
variance with the data, cf.\ the discussion in Ref.~\cite{[Shi14a]}.

\section{Conclusions}
\label{sec4}

In this work we have used three different EDF approaches, each for
two different parameter sets, so as to gain some insight into the
degree of systematic uncertainties that are related to applying these
approaches to spectroscopic properties of heavy nuclei at the gateway
to superheavy region. On the one hand, we concluded that the overall
coarse description of several spectroscopic properties of nuclei in
this region is, in general, correct. On the other hand, we identified
numerous smaller or larger differences between the results obtained
within these different EDFs. In particular, none of the studied
global EDF parameterisations precisely describes variations of
the shell structure seen in experiment. This can be associated with
small deficiencies of the obtained deformed shell properties.

On the scale of very precise spectroscopic experimental data,
differences between various EDF approaches are large. They are of the
same order as the degree of differences with experiment. This points
to still fairly large systematic uncertainties inherent to the models
currently in use. Moreover, none of the studied models could at
present be identified as the one that systematically performs
significantly better than another one. \AnA{However, it is interesting
that on the average the accuracy of the description of the excitation
energies  of deformed one-quasiparticle states in the nobelium region
is substantially better than the accuracy of the description of the
energies of dominant single-particle states at spherical shape in
odd-mass nuclei neighbouring to doubly magic $^{208}$Pb (even as compared
with the calculations which include particle-vibration coupling
\cite{AS.11,[Tar14]}.}

The obtained results suggest that further work on improving the
performance of the EDF methods is very much required. First, one can
hope that within the existing forms of EDFs, one can still find
better global parametrizations. For Skyrme EDFs, this route has
already been explored and a negative conclusion was
reached~\cite{[Kor14]}, but for covariant and Gogny EDFs similar
work has not yet been performed. Second, one can hope that various
beyond-mean-field corrections, not included in the present analysis,
may have strong impact on the results, and thus modify the current
conclusions.
At least that happens in the covariant framework, where
accounting for the (quasi)particle-vibration coupling improves
substantially the accuracy of the description of predominantly
single-particle states in spherical medium and heavy
nuclei \cite{LA.11,[Lit12]}.
In addition, there is still a possibility of building new
functionals, with beyond-mean-field effects incorporated
from the very beginning. Again, for the Skyrme EDFs, employed
together with odd-particle polarization effects included, a negative
conclusion has recently been reached~\cite{[Tar14]}.
However, so far in the covariant
and Gogny EDF approaches no such studies have been performed, and such route has to be explored.
Third, one can
also attempt building new classes of EDFs, where systematic
expansions within different schemes are used. Only very recently this
route was started to be explored in non-relativistic EDFs, see,
Refs.~\cite{[Car10e],[Sto10],[Geb11],[Dob12a],[Sad13b],[Ben14],[Rai14],[Dug15],[Dug15a]},
so its eventual impact on the physics discussed here is not yet known.

\section{Acknowledgements}
\label{sec5}

This material is based upon work supported
by the U.S. Department of Energy,
Office of Science, Office of Nuclear Physics under
Award Nos.\ DE-SC0013037 and DE-SC0008511 (NUCLEI SciDAC Collaboration),
by the NNSA's Stewardship Science Academic Alliances Program under Award
No.\ DOE-DE-NA0002574,
by the Academy of Finland and University of Jyv\"askyl\"a within the FIDIPRO
program, by the Polish National Science Center under Contract No.\
2012/07/B/ST2/03907, by the ERANET-NuPNET grant SARFEN of the Polish National
Centre for Research and Development (NCBiR),
by the CNRS/IN2P3 through PICS No.\ 6949,
by the MINECO grants Nos.\ FPA2012-34694 and FIS2012-34479,
and
by the Consolider-In\-ge\-nio 2010 program MULTIDARK CSD2009-00064.
We acknowledge the CSC-IT Center for Science Ltd., Finland, for the allocation
of computational resources.
Computations were also performed using HPC resources from the MCIA
(M{\'e}socentre de Calcul Intensif Aquitain) of the Universit{\'e} de Bordeaux
and of the Universit{\'e} de Pau et des Pays de l'Adour.

\bibliographystyle{elsarticle-num}





\end{document}